%% file: main.tex
\RequirePackage{fix-cm}
\documentclass[smallextended]{svjour3}
\smartqed  
\usepackage{graphicx}
\usepackage{subfig}
\usepackage{jnl_macros}
\usepackage{amsmath}
\usepackage{amsfonts}
\usepackage{amssymb}
\usepackage{natbib}
\usepackage{url}
\usepackage{multirow}
\usepackage{booktabs}

\newcommand{\imgdir}{./images}
\newcommand{\susi}{SUSI}
\newcommand{\pavo}{PAVO}
\newcommand{\musca}{MUSCA}
\newcommand{\mata}[1]{\mathbf{#1}}
\newcommand{\mats}[1]{\boldsymbol{#1}}
\newcommand{\ift}[1]{\mathcal{FT}^{-1}\left\{#1\right\}}
\newcommand{\ftf}[1]{\mathcal{FT}\left\{#1\right\}}
\newcommand{\fts}[1]{\hat{#1}}

\journalname{Experimental Astronomy}

\begin{document}

\title{Simulating a dual beam combiner at \susi\ for narrow-angle astrometry}
\author{Yitping Kok \and
        Vicente Maestro \and
	Michael J. Ireland \and
	Peter G. Tuthill \and
	J. Gordon Robertson
}
\authorrunning{Kok et al.}
\institute{Y. Kok, V. Maestro, P. G. Tuthill, J. G. Robertson \at
              Sydney Institute for Astronomy, \\
	      School of Physics, University of Physics, \\
	      NSW 2006, Australia. \\
              \email{y.kok@physics.usyd.edu.au}
           \and
           M. J. Ireland \at
              Department of Physics and Astronomy, \\
              Macquarie University, \\
	      NSW 2109, Australia. \\
}

\date{Received: date / Accepted: date}
\maketitle

\begin{abstract}
The Sydney University Stellar Interferometer (\susi) has two beam combiners,
i.e. the Precision Astronomical Visible Observations (\pavo) and the
Microarcsecond University of Sydney Companion Astrometry (\musca). The primary
beam combiner, \pavo, can be operated independently and is typically used to
measure properties of binary stars of less than 50 milliarcsec (mas) separation
and the angular diameters of single stars. On the other hand, \musca\ was
recently installed and must be used in tandem with the former. It is dedicated
for microarcsecond precision narrow-angle astrometry of close binary stars.  The
performance evaluation and development of the data reduction pipeline for the
new setup was assisted by an in-house computer simulation tool developed for
this and related purposes. This paper describes the framework of the simulation
tool, simulations carried out to evaluate the performance of each beam combiner
and the expected astrometric precision of the dual beam combiner setup, both at
SUSI and possible future sites.


\end{abstract}

\input{intro.tex}

\input{framework.tex}

\input{pavosim.tex}

\input{muscasim.tex}

\input{intensity2nphotons.tex}

\input{effatm.tex}

\input{testcases.tex}

\section{Conclusion}
The simulators developed in this work not only have been a very useful tool to
test the data reduction pipeline developed for the new setup and observation
technique at \susi\ but have also shown the feasibility of visible wavelength
phase-referencing at sub-wavelength OPD uncertainty. The \pavo--\musca\ setup at
\susi\ can determine the position of a fringe packet of a 2-4th magnitude star
to an accuracy of 5nm by coherently integrating 1000-2000 good fringe packet
scans. The simulated performance of the dual beam combiner can be extrapolated
to estimate performance of a similar setup at future possible sites (e.g.\ NPOI
and Antarctica). In addition to that, due to the selection of input parameters
the simulators were designed to accept they could also be used to perform
simulation for many other functions (e.g.\ to explore the option of expanding
the capability of \pavo\ at CHARA from a 3-telescope to a 4-telescope beam
combiner or to investigate the effect of optical aberration of lenses on the
performance of the beam combiners) which is beyond its main role described in
this paper. The IDL code for the simulators can be obtained from the
corresponding author via email.

\begin{acknowledgements}
Y.K. would like to acknowledge the support from the University of Sydney
International Scholarship (USydIS).
\end{acknowledgements}

\appendix
\bibliographystyle{spbasic}

\input{main.bbl}
\end{document}

%% file: intro.tex
\section{Introduction}

A dual beam combiner setup was recently installed in \susi. The main role of the
new setup is to perform high precision narrow-angle astrometry of close binary
stars. The relative position of one star on the celestial sphere with respect to
another in a binary system can be determined by measuring the separation
(in optical delay) of their fringe packets\footnote{a fringe packet is an
interference pattern produced by a light source of finite bandwidth (e.g.\
starlight) whereby the fringe visibility diminishes quickly to zero as the
optical path difference (OPD) producing the fringes deviates away from zero.}
formed by an optical long baseline interferometer like \susi. The accuracy of
the projected separation of the binary star systems obtained from this method is
determined by the uncertainty of the optical delay measurement. A more accurate
measurement of the optical delay of a fringe packet can be made by measuring the
phase delay of the fringes instead of the group delay \citep{Lawson:2000}.

Now, if the measurements are to be carried out from the ground then the position
of the pair of fringe packets must first be stabilized because their positions
are not static but constantly changing due to atmospheric turbulence. This can
be achieved using a technique called phase-referencing (PR)
\citep{Colavita:1992, Shao:1992}. In this technique, two beam combiners are
required. The phase delay of one fringe packet is measured accurately in the
presence of atmospheric turbulence with one beam combiner (usually called the
fringe tracker) and then fed-forward into another companion beam combiner to
stabilize the position of the same or another fringe packet. This technique was
demonstrated with PHASES \citep{Muterspaugh:2010a} at PTI \citep{Colavita:1999}
in the near infra-red wavelengths which had achieved an astrometric precision of
35$\mu$as (with separation less than 1$''$ close binaries) within 70 minutes of
observation time \citep{Lane:2004}.

The dual beam combiner setup in \susi\ is specifically designed to do the same
(phase-referencing observations). The main beam combiner at \susi, namely
\pavo\, is used as the fringe tracker to measure the phase delay of the fringes
of the primary star in the binary system in real time and the companion beam
combiner, namely \musca, is used to simultaneously record either the fringes of
the primary or the secondary star. This setup is similar to PHASES where both
beam combiners receive the same pair of starlight beams from the siderostats and
observe the same field of view ($<2''$) of the sky. However in many other ways
it is different. Firstly, \pavo\ and \musca\ operate in the visible wavelengths.
Secondly, each beam combiner operates at a slightly different bandwidth compared
to the other. Thirdly, the phase-referencing of stellar fringes are carried out
in post-processing which eliminates the need for a feedback servo loop in
\musca. Lastly, \musca\ observes only one stellar fringe packet at a time but
can switch between a pair of fringe packets of a binary star during observation.

With the introduction of \musca\ into \susi, the existing data reduction
pipeline was also upgraded to support the dual beam combiner configuration. The
software development, which mainly involved putting in additional features to
estimate phase delay of stellar fringes and to carry out a non-real time
phase-referencing operation, was greatly assisted by an in-house computer
simulation framework.
The framework and its usage are described in this paper.
Firstly, Section~\ref{sec:framework} gives a general overview of the simulation
framework. Subsequently, Sections~\ref{sec:pavosim}, \ref{sec:muscasim} and
\ref{sec:sim_nphotons} describe models of fringes employed by the simulators to
generate the test data sets while Section~\ref{sec:sim_effatm} describes a
method that was used to include the effect of atmospheric turbulence in the
simulation. Lastly Section~\ref{sec:sim_testcases} shows the output of several
dual beam combiner simulations and the expected performance of the instruments.

%% file: framework.tex
\section{Simulators and the framework} \label{sec:framework}

\begin{figure}
\centering
\includegraphics[width=\textwidth]{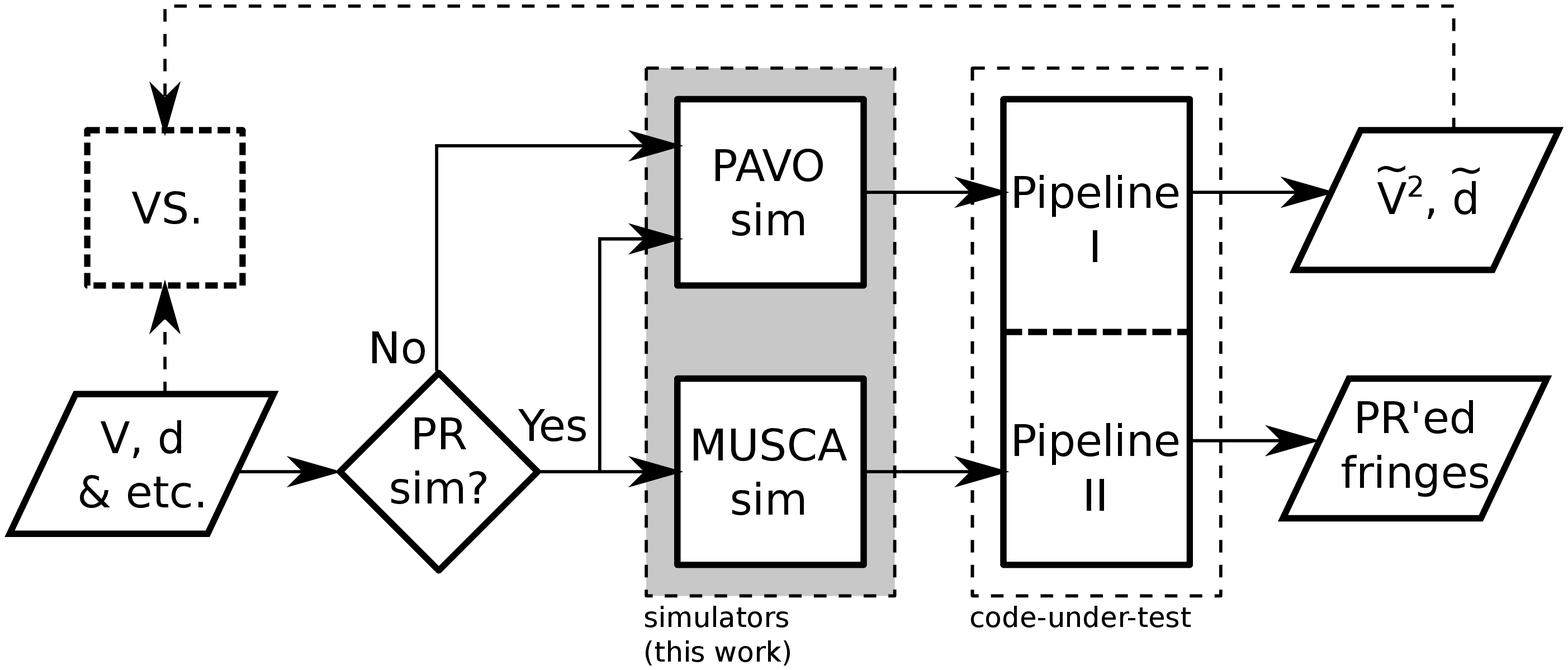}
\caption{Logical flow diagram of the data reduction pipeline test bench.}
\label{fig:framework}
\end{figure}

Two simulators, which are computer models of the \pavo\ and \musca\ beam
combiners, are developed to generate a set of simulated interferograms of each
beam combiner based on user-specified inputs. The simulators were written in the
Interactive Data Language (IDL) but the design concepts and algorithms described
here can be implemented in any other languages. Both simulators can read the
same set of user-specified inputs and by doing so allow users to simulate a dual
beam combiner operation in which stellar fringes are recorded by the actual
instruments simultaneously in real time. The simulated interferograms can
then be used to test the fringe visibility squared ($V^2$) estimation and
the phase-referencing algorithm of the upgraded data reduction pipeline
\citep{Kok:2012}. By comparing the user-specified input and the simulator
output, especially the estimated $V^2$ since it is the main science observable
of \pavo, the accurateness of the estimation and the performance of the dual
beam combiner setup can be assessed. Fig.~\ref{fig:framework} illustrates the
data reduction pipeline test bench which shows the usage of the two simulators
developed in this work.

\begin{figure}
\centering
\includegraphics[width=\textwidth]{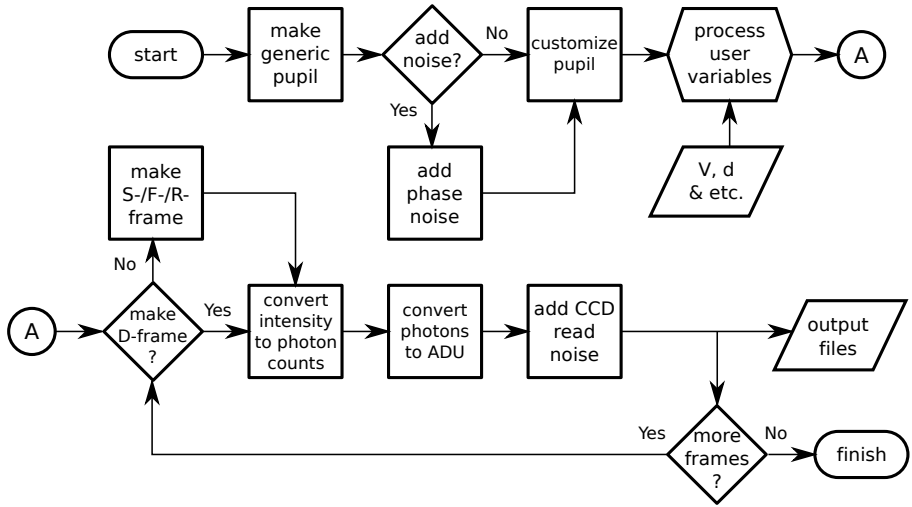}
\caption{Logical flow diagram of the PAVO and MUSCA simulators.}
\label{fig:simulator}
\end{figure}

An overall logical flow of both simulators is illustrated in
Fig.~\ref{fig:simulator}. Simulation begins with a generic model of two or more
pupils. They are then customized according to the optics of individual beam
combiner. Additional and optional phase noise to simulate the effect of
atmospheric turbulence can be included before the customization of the pupils.
Then user-specified inputs are processed, e.g.\ to determine the number of
interferograms (referred to as frames in the flow diagram) to be generated or
the visibility and phase delay of the fringes to be simulated. After generating
the required frame either by coherent or incoherent combination of the pupils,
depending on the type of frame (refer Section~\ref{sec:pavo_frames}), e.g.\
science (S-), foreground (F-), ratio (R-) or dark (D-) frame, the amplitude of
the combined pupil is converted into photon counts and subsequently into the
detector read-out units (ADU). The detector read-out noise is also included into
the simulator output. The simulated data is then saved into a file of
appropriate format. The simulation finishes when all the required frames are
generated. The details of each stage of the logical flow are discussed in the
subsequent sections.

%% file: pavosim.tex
\section{The \pavo\ simulator} \label{sec:pavosim}

\begin{figure}
\centering
\includegraphics[width=\textwidth]{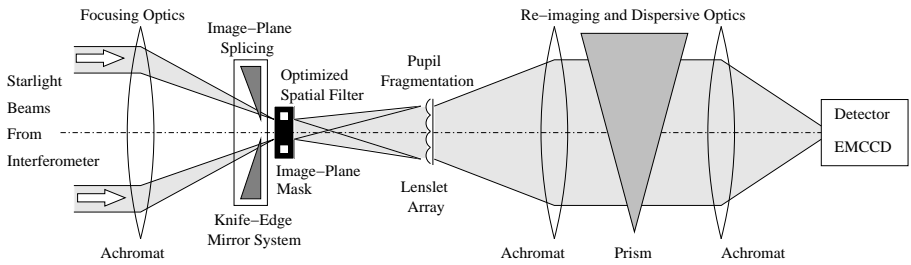}
\caption{Schematic diagram of the \pavo\ beam combiner at \susi.}
\label{fig:pavosim_sch}
\end{figure}

The \pavo\ beam combiner is a multi-axially aligned Fizeau-type interferometer.
But unlike a typical Fizeau interferometer, \pavo\ forms spatially modulated
interference fringes in the pupil plane of the interferometer and then
spectrally disperses the fringes with an integral field unit. It also employs
spatial filtering in its image plane and an array of cylindrical lenslet to
utilize the full multi-r$_0$ aperture of the siderostats at \susi.
The lenslet array fragments the pupil of the siderostats into several segments
so that fringes from different parts of the pupil can be measured separately.
The schematic diagram of
the \pavo\ beam combiner is shown in Fig.~\ref{fig:pavosim_sch}. It combines
starlight beams from any 2 of the 11 siderostats \citep{Davis:1999} at one time.
Despite that, the \pavo\ simulator developed in this work is able to simulate
higher order beam combination (i.e.\ 3 and more beams simultaneously) because
the optical design of \pavo\ at \susi\ is an adaptation of a twin instrument at
the CHARA array on Mount Wilson, California \citep{McAlister:2005}, which has
the capability of combining starlight beams from up to 3 telescopes. The
original optical design of \pavo\ at CHARA and the modified version at \susi\
have been discussed in detail by \citet{Ireland:2008} and \citet{Robertson:2010}
respectively.

\subsection{Input and output of the simulator}
\begin{figure}
\centering
\subfloat[]{\includegraphics[width=\textwidth]{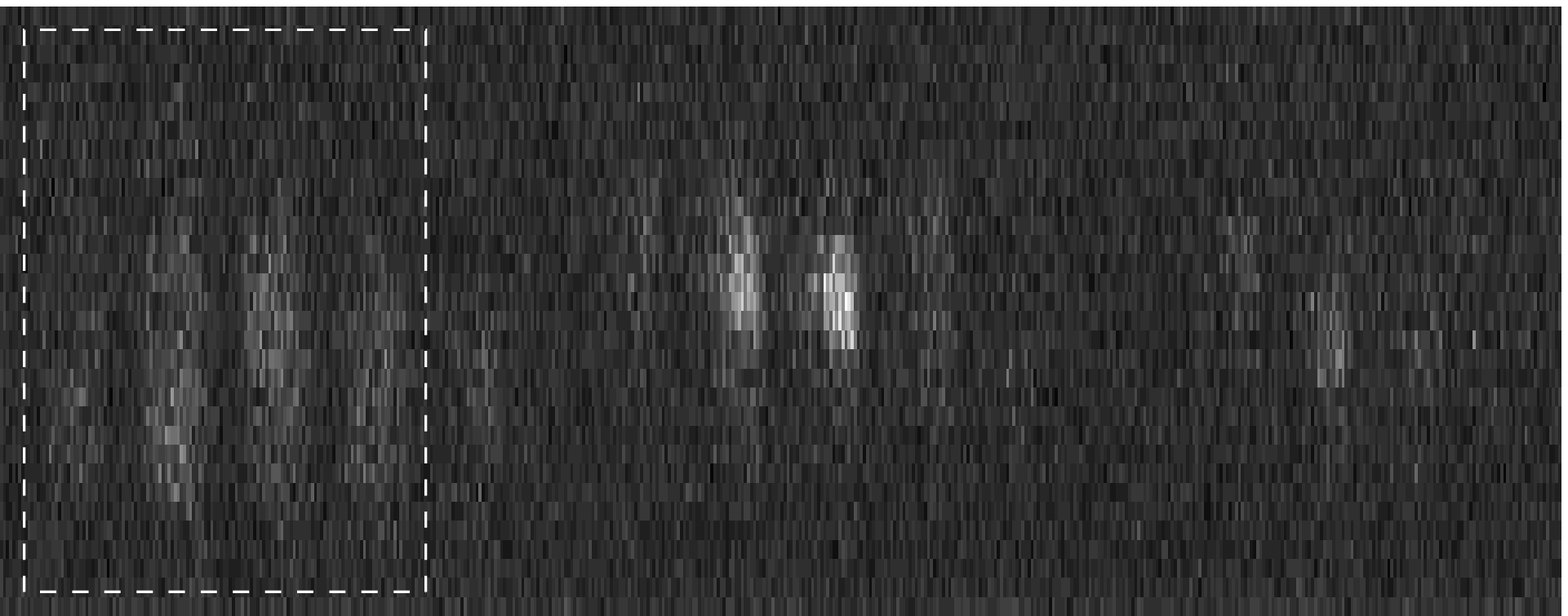}}\\
\vspace{-0.5em}
\subfloat[]{\includegraphics[width=0.49\textwidth]{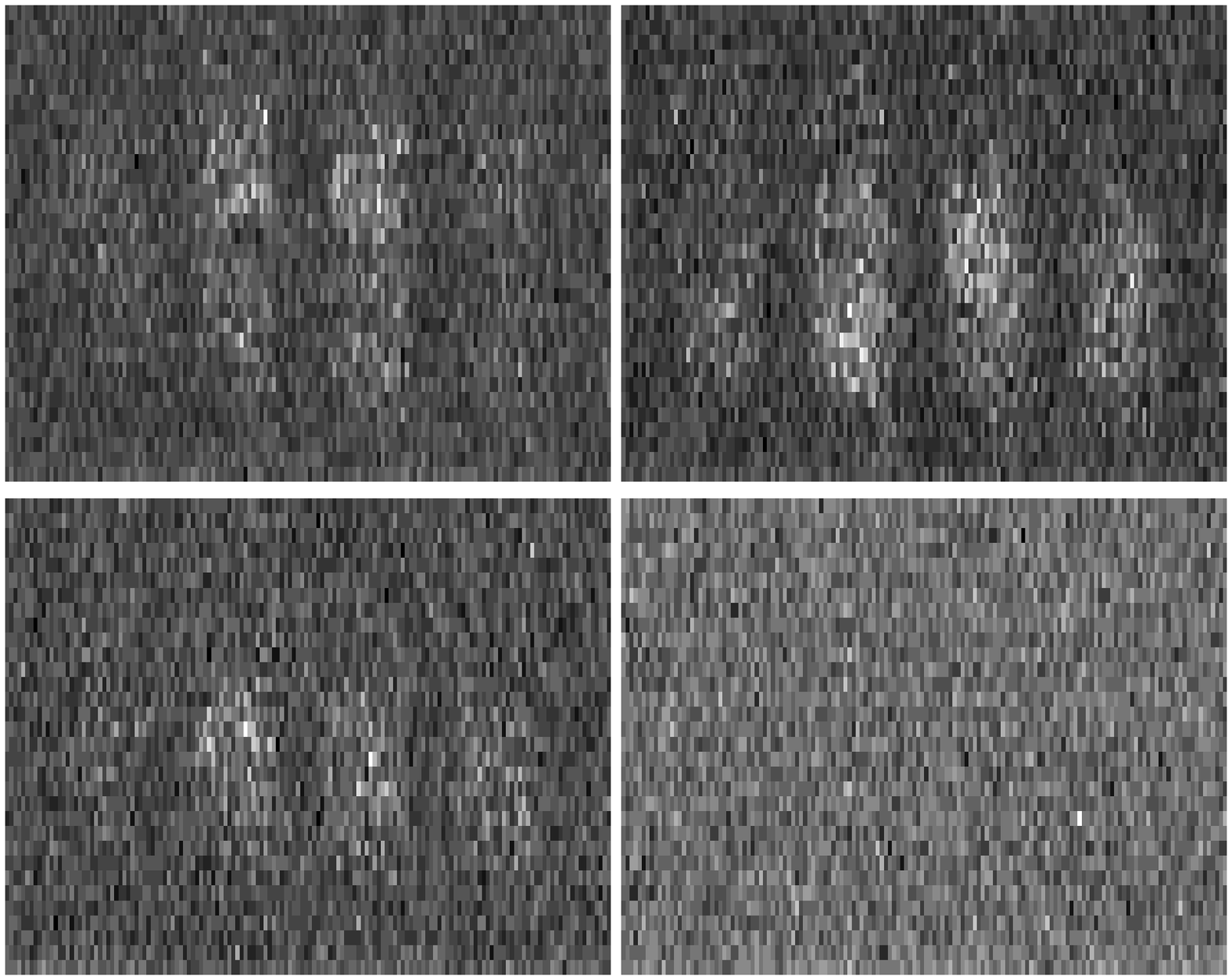}}
\hspace{0.5em}
\subfloat[]{\includegraphics[width=0.49\textwidth]{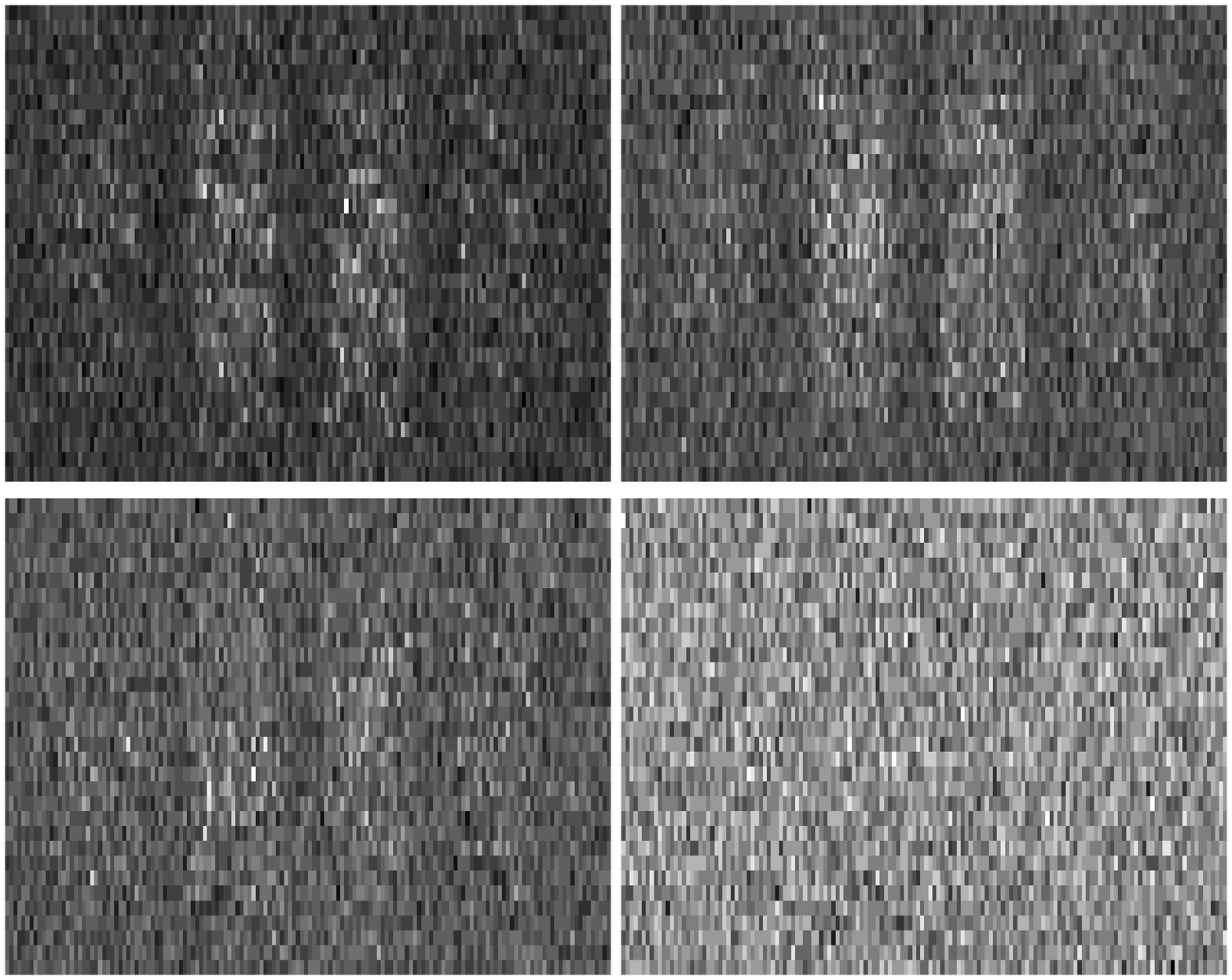}}\\
\caption{(a) An example of real on-sky data of $\alpha$ Gru recorded by the \pavo\
camera. It contains images of spectrally dispersed (horizontally) pupils as sampled by
an array of four lenslets. The science pupil on the left (indicated by the dotted box)
of the frame is where fringes are formed by a pair of overlapped pupils. The pupils in
the middle and left of the frame are used for tip-tilt control. They are ignored and will
not be generated during simulation. The images of the science pupil for different frame
types are shown in (b) and (c). The images, from left to right and top to bottom, belong
to four different types of frames, namely science, foreground, ratio and dark
respectively. The images in (b) are recorded by the camera while those in (c) are
generated by the simulator. The convention adopted by the mathematics in this paper
takes the \emph{vertical} axis, which is the fringes direction, as the $x$-axis and the
\emph{horizontal} axis, which is the spectral channel direction, as the $y$-axis.}
\label{fig:pavosim_fits}
\end{figure}

%
%
There are more than 20 input parameters to the \pavo\ simulator
with the more critical ones
listed in Table~\ref{tab:pavosim_inputs}. Some input parameters are common to the
\musca\ simulator which will be described in Section~\ref{sec:muscasim}.
The main output of the simulator is a set of
FITS\footnote{\url{http://fits.gsfc.nasa.gov}} formatted files which contain images of
simulated interferograms, hereinafter referred to as frames, as well as header
information (e.g.\ timestamp, fringe lock status, etc) as would be recorded by the
\pavo\ camera.  There are four different types of frames recorded by the camera during
actual observations. There are the science, ratio, foreground and dark frames. Examples
of each of these frames are shown in Fig.~\ref{fig:pavosim_fits}. Each frame, except the
dark, contains images of the spectrally dispersed (horizontally in the figure) pupil as
sampled by an array of lenslets. The number of lenslets, $N_{\rm{LL}}$, is different
between PAVO at SUSI and at CHARA. The example shown in Fig.~\ref{fig:pavosim_fits} is
of the former which has the lenslets arranged in a one-dimensional array and
there are 4 lenslets per pupil. The left pupil is for science while the middle
and right pupils are used for tip-tilt correction and therefore are ignored in
the simulation. Frames of PAVO at CHARA contains only the science pupil.

\begin{table}
\center
\caption{Input parameters for the \pavo\ (P) and \musca\ (M) simulators}
\label{tab:pavosim_inputs}
\begin{tabular}{c c l}
\hline
Name & Simulators & Description \\
\hline
$t_{\rm{START}}$	& P,M	& Start of simulation in Julian date \\
$t_{\rm{STEP}}$		& P,M	& Exposure time of camera/photodetector \\
$N_{\rm{S-FITS}}$	& P	& Number of science type FITS to generate \\
$N_{\rm{R-FITS}}$	& P	& Number of ratio type FITS to generate \\
$N_{\rm{F-FITS}}$	& P	& Number of foreground type FITS to generate \\
$N_{\rm{D-FITS}}$	& P	& Number of dark type FITS to generate \\
$N_{\rm{MED}}$		& P,M	& Number and types of optical media \\
$\mats{\zeta}$		& P,M	& Astrometric OPD in m \\
$N_{\rm{TEL}}$		& P,M	& Number of telescopes \\
$\mata{B}$		& P,M	& Details of telescopes (e.g. baselines) to be used \\
$m_V$			& P,M	& Magnitude of source in V band \\
$\mata{V}$		& P,M	& Model complex visibility of source \\
$\mata{D},\mata{d}$	& P,M	& Offset of $\mats{\zeta}$ in m \\
$r_0$			& P,M	& Fried parameter in m \\
$\sigma_{r_0}^{-1}$	& P,M	& Wavelength in which $r_0$ is specified ($\mu$m) \\
$\tau_0$		& P,M	& Coherence time in milliseconds \\
$L_0$			& P,M	& Outer scale of atmospheric turbulence in m \\
$t_{\rm{STEP}}$		& M	& Time interval between steps \\
$N_{\rm{STEP}}$		& M	& Number of steps per scan \\
$N_{\rm{SCAN}}$		& M	& Number of scan to simulate \\
$L_{\rm{SCAN}}$		& M	& Length of a scan in $\mu$m \\
\hline
\end{tabular}
\end{table}

\subsection{Model pupil} \label{sec:pavosim_optics}

Before the pupils are combined to form
fringes, they are first spatially filtered in the image plane with square apertures
(hereinafter referred to as the \pavo\ mask), one for each beam, to remove high spatial
frequency noise arising as a consequence of atmospheric turbulence and optical
aberration. A pupil from a telescope is modeled by a square matrix, $\tilde{\mata{P}}$,
of size $N_{\rm{FFT}} \times N_{\rm{FFT}}$,
\begin{equation} \label{eq:pavosim_pupil}
\begin{split}
\tilde{\mata{P}} &= \left\{
  \begin{array}{l l}
  \mata{J_{N_{\rm{FFT}},N_{\rm{FFT}}}} & \quad;\text{perfect wavefront} \\
  \mata{J_{N_{\rm{FFT}},N_{\rm{FFT}}}} \circ
  \exp\left(i\mats{\varphi}\right) & \quad;\text{corrugated wavefront} \\
  \end{array}\right. \\
\end{split}\end{equation}
where $\mata{J}$ represents a unit matrix consisting of all 1s of size
(row$\times$column) indicated by its subscript and the phase component is either
zero or
$\mats{\varphi}$ so as to represent a perfect or corrugated wavefront. A corrugated
wavefront due to atmospheric turbulence will be elaborated in Section~\ref{sec:sim_effatm}.
The notation $\circ$ is the Hadamard or the element-by-element multiplication
operator of two matrices. The model of a spatially filtered pupil, $\mata{P'}$, is then,
\begin{equation} \begin{split}
\mata{P'} &= \ift{\Box \circ \ftf{\bigcirc\circ\tilde{\mata{P}}}} \\
\end{split}\end{equation}
where $\bigcirc$ and $\Box$ are square matrices of the same size which define a circular
pupil and the spatial filter respectively. Each element in the matrices is defined as,
\begin{equation} \label{eq:pavosim_circbox}
\begin{split}
\bigcirc_{u,v} &= \left\{
  \begin{array}{l l}
  1 & \quad;\text{if}\;\sqrt{(u-N_{\rm{FFT}}/2)^2+(v-N_{\rm{FFT}}/2)^2)} \leq N_{\rm{DIA}}/2 \\
  0 & \quad;\text{otherwise} \\
  \end{array}\right. \\
\Box_{x,y} &= \left\{
  \begin{array}{l l}
  1 & \quad;\text{if}\;|x-N_{\rm{FFT}}/2| \leq N_{\rm{MASK}}/2\;\text{and}\;|y-N_{\rm{FFT}}/2| \leq N_{\rm{MASK}}/2 \\
  0 & \quad;\text{otherwise} \\
  \end{array}\right. \\
\end{split}\end{equation}
where $N_{\rm{DIA}}$ and $N_{\rm{MASK}}$ are relative sizes of the diameter of the
telescopes and the width of the \pavo\ mask respectively. $N_{\rm{FFT}}$, $N_{\rm{DIA}}$
and $N_{\rm{MASK}}$ take only integer values.
The ratio of the width of the mask to the size of the image varies with the wavelength
of light and is defined in Eq.~\eqref{eq:pavosim_nmask}. In this paper the wavelength of
light is always represented by its reciprocal, or wavenumber, $\sigma$ or $\sigma_j$,
where $j$ is the index of vector $\mats{\sigma}$ which represents a range of
wavenumbers applicable to
\pavo. The wavelength scaling factor in Eq.~\eqref{eq:pavosim_nmask} is derived based on the
$f$-ratio of the beam and the distance of the \pavo\ mask from the pupil plane. The
different values between \pavo\ at \susi\ and CHARA means the optical setup at the two
interferometers are not exactly the same. The simulator keeps the denominator of the
left-hand side (LHS) of Eq.~\eqref{eq:pavosim_nmask} constant and varies $N_{\rm{MASK}}$
at different wavelengths of light to satisfy the equation.
\begin{equation} \label{eq:pavosim_nmask} \begin{split}
\frac{N_{\rm{MASK}}}{N_{\rm{FFT}}/N_{\rm{DIA}}} &\approx \left\{
  \begin{array}{l l}
  1.7\sigma_j & \quad;\text{\susi} \\
  8.5\sigma_j & \quad;\text{CHARA} \\
  \end{array}\right. \\
\end{split}\end{equation}
With $N_{\rm{FFT}}$ and $N_{\rm{DIA}}$ typically set to 256 and 114
respectively, the range of values of $N_{\rm{MASK}}$ is 5--7 for \susi\ and
22--30 for CHARA.

Before a model pupil is used to form fringes, it is down-sampled to the size of an
actual image taken by the \pavo\ camera, which is $N_{\rm{X}} \times N_{\rm{LL}}$, where
$N_{\rm{X}}$ is the number of pixels in the $x$-axis and $N_{\rm{LL}}$ is the number of
lenslets in the $y$-axis. The values of each parameter are listed in
Table~\ref{tab:pavosim_params}. As a result, the model pupil becomes,
\begin{equation} \label{eq:pavosim_pupil_small} \begin{split}
\mata{P} &=
  \frac{1}{N_{\rm{X}}N_{\rm{LL}}}
  \mats{\mathcal{I}}\mata{^T_{N_{\rm{FFT}},N_{\rm{X}}}}\;
  \mata{P'}\;
  \mats{\mathcal{I}}\mata{_{N_{\rm{FFT}},N_{\rm{LL}}}}\;
\end{split}\end{equation}
where $\mats{\mathcal{I}}\mata{_{M,N}}$ is a M$\times$N matrix,
\begin{equation} \begin{split}
\mats{\mathcal{I}}\mata{_{M,N}} &=
\begin{bmatrix}
\mata{J_{M/N,1}} & 0 & \cdots & 0 \\
0 & \mata{J_{M/N,1}} & \cdots & 0 \\
\vdots & \vdots & \ddots & \vdots \\
0 & 0 & \cdots & \mata{J_{M/N,1}} \\
\end{bmatrix}
\end{split}\end{equation}
and $\mata{J_{M/N,1}}$ is a $M/N\times1$ unit matrix consisting of all 1s.
This simple down-sampling only works if $M$ is an integer multiple of
$N$. For example,
to down-sample a 6$\times$6 matrix $\mata{P}'$ to a 2$\times$3 matrix $\mata{P}$, the following operation
can be applied,
\begin{equation} \begin{split}
\mata{P} &= \frac{1}{6}
\begin{bmatrix}
1 & 0 \\
1 & 0 \\
1 & 0 \\
0 & 1 \\
0 & 1 \\
0 & 1 \\
\end{bmatrix}^{\mata{T}}
\begin{bmatrix}
P_{1,1}' & P_{1,2}' & \cdots & P_{1,6}' \\
P_{2,1}' & P_{2,2}' & \cdots & P_{2,6}' \\
\vspace{-0.4em}\\
\vdots & \vdots & \ddots & \vdots \\
\vspace{-0.4em}\\
P_{6,1}' & P_{6,2}' & \cdots & P_{6,6}' \\
\end{bmatrix}
\begin{bmatrix}
1 & 0 & 0 \\
1 & 0 & 0 \\
0 & 1 & 0 \\
0 & 1 & 0 \\
0 & 0 & 1 \\
0 & 0 & 1 \\
\end{bmatrix}
\end{split}\end{equation}
where the first and the last matrices on the right-hand side (RHS) are
$\mats{\mathcal{I}}\mata{_{6,2}}$ and $\mats{\mathcal{I}}\mata{_{6,3}}$ respectively.

The values of the physical parameters in Eq.~\eqref{eq:pavosim_nmask},
Eq.~\eqref{eq:pavosim_pupil_small} and other equations in this section are listed in
Table~\ref{tab:pavosim_params}.
\begin{table}
\center
\caption{Specification of \pavo}
\label{tab:pavosim_params}
\begin{tabular}{l c c c}
\hline
Parameter & Notation & \pavo@CHARA & \pavo@\susi \\
\hline
Number of spectral channels	& $N_{\rm{\sigma}}$ 	& 19		& 21 \\
Spectral range ($\mu$m$^{-1}$)	& $\mats{\sigma}$	& 1/0.88 -- 1/0.63& 1/0.80 -- 1/0.53  \\
Number of lenslets		& $N_{\rm{LL}}$		& 16		& 4 \\
Number of pixels in x-axis	& $N_{\rm{X}}$		& 128		& 32 \\
FOV to pupil size ratio		& $\tilde{R}$		& 5.8		& 1.8 \\
\hline
\end{tabular}
\end{table}

\subsection{Simulating various types of frames} \label{sec:pavo_frames}

Different types of frames are generated using different combinations of model pupils
described in Section~\ref{sec:pavosim_optics}. Sets of 50 frames are saved into a FITS
file. Each FITS file has a timestamp of, 
\begin{equation} \label{eq:pavosim_timestamp}
t_\alpha = t_{\rm{START}} + 50\times \alpha t_{\rm{STEP}}
\end{equation}
where $\alpha$ is the number of FITS files already generated before the current one. The
typical value of $t_{\rm{STEP}}$ for \pavo\ at SUSI and CHARA is 5ms and 8ms
respectively. The number of files to be generated for each type of frame is determined
by the user.

\subsubsection{Science frames}

The science frames are generated using two or three pupils, depending on the number of
telescopes in use. Simulation of \pavo\ at CHARA can use up to three. In reality the
pupils are aligned to overlap each other and combined to produce spatially modulated
fringes across the pupils.
The model intensity across the overlapped pupils at one particular wavelength is given
as,
\begin{equation} \label{eq:pavosim_F}
\begin{split}
\mata{F} &= w_j\sum_{\theta=1}^{N_{\rm{TEL}}} \sum_{\tilde{\theta}=1}^{N_{\rm{TEL}}}
  V_{\theta,\tilde{\theta}}\,
  a_\theta\mata{P}^{(\theta)} \circ a_{\tilde{\theta}}\overline{\mata{P}}^{(\tilde{\theta})}\,
  \circ \exp\left(i\mata{\Phi}^{(\theta,\tilde{\theta})}\right) \\
\end{split}\end{equation}
where the notation $\overline{\mata{P}}$ represents the complex conjugate of the
variable $\mata{P}$. The indices $\theta$ and $\tilde{\theta}$ denote one of the several
pairs of telescopes
used in the simulation ($N_{\rm{TEL}}$) while $a_\theta$
denotes the weighted amplitude of $\mata{P}^{(\theta)}$ such that,
\begin{equation} \label{eq:pavosim_a}
\begin{split}
a_1^2 + a_2^2 + \cdots + a_{N_{\rm{TEL}}}^2 = 1
\end{split}\end{equation}
This condition has no physical reason but is imposed for the convenience of scaling the
normalize intensity to the right photon rate in Section~\ref{sec:sim_nphotons}.
The term $w_j$ states the relative intensity of the summation in
Eq.~\eqref{eq:pavosim_F} at one wavelength while the vector $\mata{w}$ which $w_j$ is a part
of describes the spectrum of the light source and the bandpass profile of \pavo,
\begin{equation} \label{eq:pavosim_w}
\begin{split}
\mata{w} = 
\begin{bmatrix}
w_1 & w_2 & \cdots & w_{N_{\sigma}}
\end{bmatrix}\;\;
;\text{where } \mata{wJ_{N_{\sigma},1}} = 1
\end{split}\end{equation}
This term can easily be customized by user according to the need of a
simulation. However, the results shown in Section~\ref{sec:sim_testcases} were
simulated with a smooth-edged top hat function for $\mata{w}$.
The model of fringes, $\mata{F}$, is expressed in this form in order to allow
a model of complex fringe visibility, $V_{\theta,\tilde{\theta}}$, to be applied to the
pairs of pupils. The complex fringe visibility matrix, which is a user supplied input,
is defined as,
\begin{equation} \label{eq:pavosim_V}
\begin{split}
\mata{V} &=
\begin{bmatrix}
1 & V(\sigma_jB_{1,2}) & \cdots & V(\sigma_jB_{1,N_{\rm{TEL}}}) \\
\overline{V}(\sigma_jB_{1,2}) & 1 & \cdots & V(\sigma_jB_{2,N_{\rm{TEL}}}) \\
\vdots & \vdots & \ddots & \vdots \\
\overline{V}(\sigma_jB_{1,N_{\rm{TEL}}}) & \overline{V}(\sigma_jB_{2,N_{\rm{TEL}}}) & \cdots & 1 \\
\end{bmatrix} \\
\end{split}\end{equation}
where each off diagonal element represents the fringe visibility of a model light source
at a given wavelength and baseline. The term $B_{\theta,\tilde{\theta}}$ is the
magnitude of a baseline vector $\vec{B}_{\theta,\tilde{\theta}}$ formed by a pair of
telescopes $\theta$ and $\tilde{\theta}$.
%
Lastly the matrix $\mata{\Phi}$
represents an additional phase difference between the two pupils and it is used to model
the difference in piston and tilt in the wavefront of the pupils. 
\begin{equation} \label{eq:pavosim_Phi}
\begin{split}
\mata{\Phi}^{(\theta,\tilde{\theta})} &= 2\pi\sigma_j\left(
  \zeta_{\theta,\tilde{\theta}} + [\mata{N}]_{j,*}\mata{z}^{(\theta,\tilde{\theta})} + D_{\theta,\tilde{\theta}} +
  N_{j,1}S_{\theta,\tilde{\theta}}\tilde{R}\mata{x}\mata{J_{1,N_{\rm{LL}}}}
  \right) \\
\end{split}\end{equation}

The first three terms in Eq.~\eqref{eq:pavosim_Phi} represent the piston term. The first
term $\zeta_{\theta,\tilde{\theta}}$ is the astrometric OPD due to the position of the
target star with respect to the baseline $B_{\theta,\tilde{\theta}}$ while the second
term $[\mata{N}]_{j,*}\mata{z}^{(\theta,\tilde{\theta})}$ is the optical path of the
delay line used to compensate the astrometric OPD.
\begin{equation} \begin{split}
\mats{\zeta} &= 
\begin{bmatrix}
0 & \hat{s}\cdot\vec{B}_{1,2} & \cdots & \hat{s}\cdot\vec{B}_{1,N_{\rm{TEL}}} \\
-\hat{s}\cdot\vec{B}_{1,2} & 0 & \cdots & \hat{s}\cdot\vec{B}_{2,N_{\rm{TEL}}} \\
\vdots & \vdots & \ddots & \vdots \\
-\hat{s}\cdot\vec{B}_{1,N_{\rm{TEL}}} & -\hat{s}\cdot\vec{B}_{2,N_{\rm{TEL}}} & \cdots & 0 
\end{bmatrix} \\
\end{split}\end{equation}
The optical delay line can comprise of various optical media. The types of optical
media are specified by the user but practically it is not more than 4 different types
(e.g.\ vacuum, air and two types of glass, BK7 or F7). The refractive
indices for each medium at the wavenumbers of \pavo\ are calculated using values and
constants obtained from \citet{Tango:1990}. The notation $[\mata{N}]_{j,*}$ represents
the $j$-th row of the refractive indices matrix, $\mata{N}$, where,
\begin{equation} \label{eq:pavosim_N} \begin{split}
\mata{N} &=
\begin{bmatrix}
n_1(\sigma_1) & n_2(\sigma_1) & \cdots & n_{N_{\rm{MED}}}(\sigma_1) \\
n_1(\sigma_2) & n_2(\sigma_2) & \cdots & n_{N_{\rm{MED}}}(\sigma_2) \\
\vdots & \vdots & \ddots & \vdots \\
n_1(\sigma_{N_\sigma}) & n_2(\sigma_{N_\sigma}) & \cdots & n_{N_{\rm{MED}}}(\sigma_{N_\sigma}) \\
\end{bmatrix}
\end{split}\end{equation}
and $n_i$ is the refractive index of one optical medium. In order to set a convention,
the first medium ($i=1$) is air.
Each element in the column vector $\mata{z}$ represents the optical path length of each
medium and to optimally compensate a given astrometric OPD the values of $\mata{z}$ are
calculated using the method described by \citet{Tango:1990}.
The third term in Eq.~\eqref{eq:pavosim_Phi} is an offset term to allow users to
simulate a non-optimally compensated astrometric OPD. The user input matrix
$\mata{D}$ is defined as,
\begin{equation} \begin{split}
\mata{D} &= 
\begin{bmatrix}
0 & d_{1,2} & \cdots & d_{1,N_{\rm{TEL}}} \\
-d_{1,2} & 0 & \cdots & d_{2,N_{\rm{TEL}}} \\
\vdots & \vdots & \ddots & \vdots \\
-d_{1,N_{\rm{TEL}}} & -d_{2,N_{\rm{TEL}}} & \cdots & 0 
\end{bmatrix} \\
\end{split}\end{equation}
where $d_{1,2}$ for example is the OPD offset for baseline $B_{1,2}$.

The last term in Eq.~\eqref{eq:pavosim_Phi} is the OPD caused by the differential wavefront
tilt between a pair of pupils at the pupil plane. It is proportional to the
separation of the apertures on the \pavo\ mask and inversely proportional to the
distance of the pupil plane from the mask.
$S_{\theta,\tilde{\theta}}$ is the ratio of the separation of the apertures on the
\pavo\ mask to the width of each aperture and is defined as,
\begin{equation} \label{eq:pavosim_S}
\begin{split}
S_{\theta,\tilde{\theta}} &= 2|\theta-\tilde{\theta}| \\
\end{split}\end{equation}
$\tilde{R}$ is the ratio of the field of view (FOV) of the \pavo\ camera to the diameter
of one pupil at $\sigma=1\mu m^{-1}$ and its value is given in
Table~\ref{tab:pavosim_params}. Lastly, $\mata{x} \in \mathbb{R}: -1/2 \leq x_{i} \leq
1/2$ is a column vector of length $N_{\rm{X}}$ which represents the pixels across the
field of
view of the camera along the direction of the tilt. This direction is also the axis
where interference fringes are formed across the camera and is referred to as the
$x$-axis by convention.

Now, $\mata{F}$ is defined at just one wavelength. In reality, the combined pupils are
spectrally dispersed by a prism. In order to model this $\mata{F}$ is evaluated
$N_{\sigma}$ times, each time at a different wavelength within the \pavo\ spectral
bandwidth. Multiple $\mata{F}$ matrices are then rearranged in the following order to
mimic the actual interferogram recorded by the camera using only their real parts (as
denoted by the notation $\Re$),
\begin{equation} \label{eq:pavosim_Fi}
\begin{split}
\mats{\mathcal{F}} &= N_{\rm{PHOTONS}} \\
  &\quad \times \Re\left\{
  \begin{bmatrix}
  \mata{0}\, F_{*,0}^{(1)} F_{*,0}^{(2)} \cdots F_{*,0}^{(N_{\sigma})}\,
  \mata{0}\, F_{*,1}^{(1)} \cdots F_{*,1}^{(N_{\sigma})}\,
  \mata{0}\, F_{*,N_{\rm{LL}}}^{(1)} \cdots F_{*,N_{\rm{LL}}}^{(N_{\sigma})}\,
  \mata{0} \\
  \end{bmatrix}
  \right\} \\
  &\quad + \mats{\epsilon}
\end{split}\end{equation}
The superscripts in parentheses represent the matrices evaluated at different
wavelengths within the spectral bandwidth. The interferogram, $\mats{\mathcal{F}}$,
which takes only the real part of the $\mata{F}$, is padded with columns of zeros,
$\mata{0}$. This is done according to a \pavo\ parameter definition file. The file
describes the pixels locations of a spectral channel within the camera's field of view
which was determined through calibration with up to two lasers. The scaling factor,
$N_{\rm{PHOTONS}}$, in Eq.~\eqref{eq:pavosim_Fi} converts intensity to energy in terms of
number of photons. This factor is proportional to the brightness of the target star,
number of telescopes used and the exposure time, all given by the user. A noise term,
$\mats{\epsilon}$, is added to the interferogram to simulate photon noise,
multiplication noise and read noise of the EMCCD camera. It is not purely an
additive term as suggested because the expression of the noise term in
Eq.~\eqref{eq:pavosim_Fi} is simplistic. Physical models are used in the
simulation to generate the photon and multiplication noise components based on
the number of photons.

\subsubsection{Foreground frames}

Foreground frames are generated in a very similar way to the science frames. Instead of
setting $\mata{D}$ to values within the coherent length of the \pavo\
spectral bandwidth, it is set to a very large number (e.g. 1m) so that no fringes are
formed across the pupils. Furthermore the visibility of the fringes are set to zero. In
the actual beam combiner, foreground frames are recorded by giving the optical delay
line a large offset from its last position where fringes were found.

\subsubsection{Ratio frames}

Ratio frames are generated using only one pupil at a time. In the actual beam combiner,
such frames are recorded when one of the many beams is blocked from reaching the camera.
This type of frame is used by the data reduction pipeline to determine the intensity of
pupil from each telescope. With only one pupil, the $\mata{F}$ matrix becomes,
\begin{equation} \begin{split}
\mata{F} &= a_\theta\mata{P}^{(\theta)} \\
\end{split}\end{equation}

\subsubsection{Dark frames}

Unlike previous types, dark frames are generated without any pupil. In the actual beam
combiner, dark frames are recorded when all the beams are blocked from reaching the
camera. The interferogram contains only the noise term which in this case made up of
only the read noise of the camera. This type of frame is used by the data reduction
pipeline to subtract the noise floor in the interferogram.
\begin{equation} \begin{split}
\mats{\mathcal{F}} &= \mats{\epsilon}
\end{split}\end{equation}

%% file: muscasim.tex
\section{The \musca\ simulator} \label{sec:muscasim}

\begin{figure}
\centering
\includegraphics[width=0.8\textwidth]{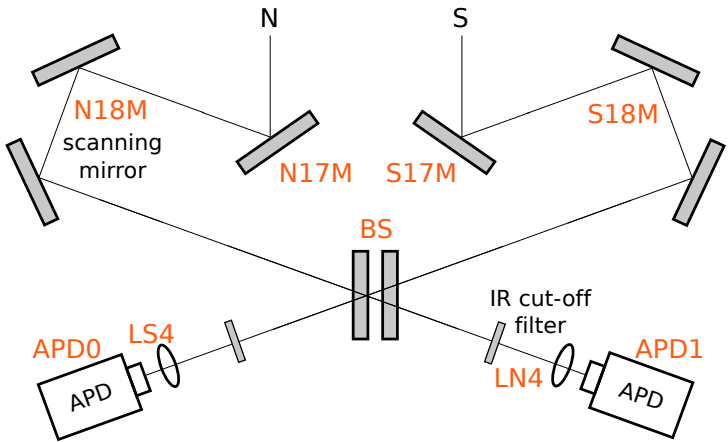}
\caption{Schematic diagram of the \musca\ beam combiner at \susi. The beamsplitter in
the diagram is labeled BS while the APDs are single pixel avalanche photodiodes used for
photon-counting.}
\label{fig:muscasim_sch}
\end{figure}

The \musca\ beam combiner is a co-axially aligned pupil-plane Michelson interferometer.
It combines only two beams at one time, each from one siderostat. There are no spatial
fringes in the image plane. If the image is diffraction limited, the Airy disk will be
completely dark when the pupils are out of phase and bright when the pupils are in
phase. \musca\ produces interference fringes by varying the difference in piston between
the two pupils through time. This is done by changing the optical path length of one
pupil in air by moving a mirror in discrete steps between two locations back and forth
rapidly. The mirror, hereinafter referred to as the scanning mirror (N18M in
Fig.~\ref{fig:muscasim_sch}), makes a scan by moving from one extreme position to
another. The physical parameters related to the scanning mirror and the operational
spectral range of \musca\ are listed in Table~\ref{tab:muscasim_params}. The schematic
diagram of the \musca\ beam combiner is shown in Fig.~\ref{fig:muscasim_sch}. Fringes
are formed at the beamsplitter (BS) and are recorded by a pair of avalanche photodiodes
(APDs), one on each sides. However the details of its optical design are discussed in
another paper \citep{Kok:2012}.

\begin{table}
\center
\caption{Specification of \musca}
\label{tab:muscasim_params}
\begin{tabular}{l c c}
\hline
Parameter & Notation & Typical values \\
\hline
Spectral range ($\mu$m$^{-1}$)	& $\mats{\sigma}$	& 1/1.0 -- 1/0.77	\\
Scan range ($\mu$m)		& $L_{\rm{SCAN}}$	& 30, 140	\\
Number of steps per scan	& $N_{\rm{STEP}}$	& 256, 1024	\\
Time interval between steps (ms)& $t_{\rm{STEP}}$	& 0.3	\\
\hline
\end{tabular}
\end{table}

\subsection{Input and output of the simulator}

%
%
%
The format of the input to the \musca\ simulator is exactly the same as the input to the
\pavo\ simulator. Some input parameters are common to both simulators but there are some
parameters which are applicable only to the \musca\ simulator. The parameters are listed
in Table~\ref{tab:pavosim_inputs}.

\begin{figure}
\centering
\subfloat[]{\includegraphics[width=0.5\textwidth]{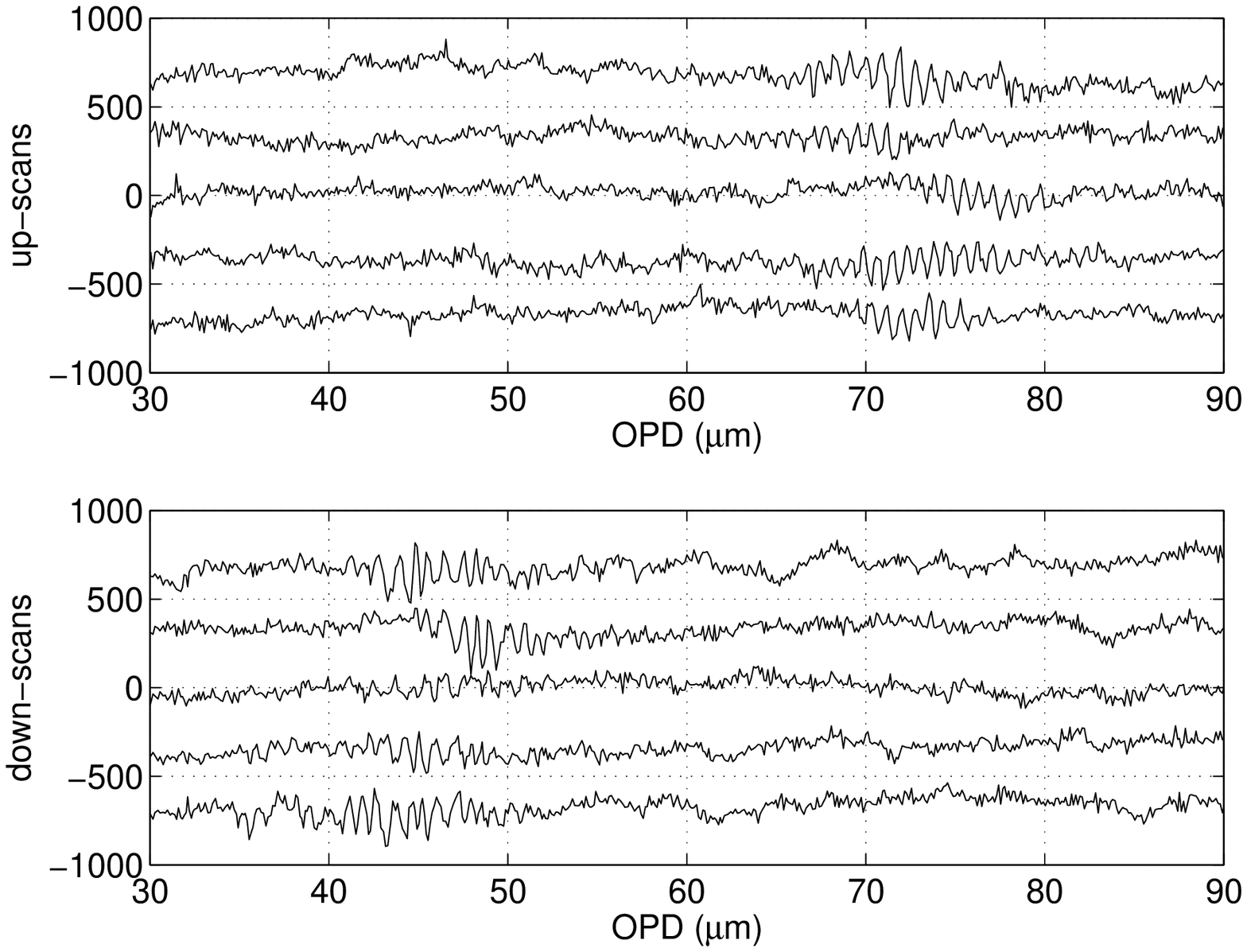}}
\subfloat[]{\includegraphics[width=0.5\textwidth]{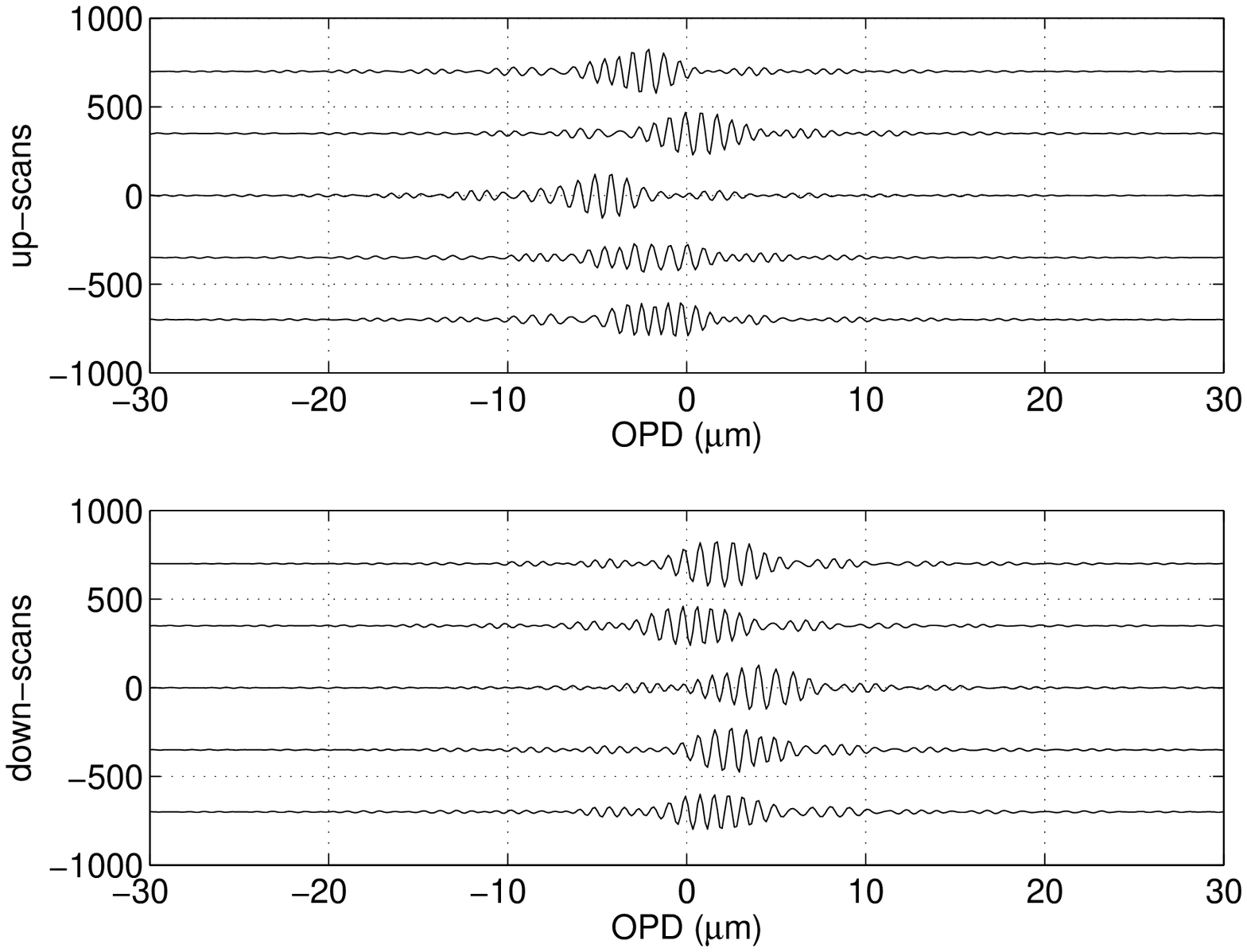}}
\caption{An example of two sets of time series of photon counts (a) recorded by the
actual MUSCA beam combiner and (b) generated by the simulator plotted as separate scans.
Each scan is plotted with an offset for visualization. The $x$-axis shows the relative
OPD within the scan range of the scanning mirror. The top and bottom plots are scans of
different direction.}
\label{fig:muscasim_txts}
\end{figure}

Instead of FITS, however, the output of the \musca\ simulator is a plain text file,
which has the same format as one generated by the actual beam combiner. The file
contains a time series of photon counts recorded by the APDs in the image plane as the
scanning mirror periodically scans through a predetermined scan range. Each photon count
has a timestamp with a precision of 10 microseconds. Fig.~\ref{fig:muscasim_txts} shows
an example of the photon counts recorded by the actual instrument as well as a set
generated by the simulator. 

\subsection{Model pupil} \label{sec:muscasim_pupil}

The model of a pupil in \musca\ is straightforward as there is no spatial filter in
\musca. Suppose the model pupil is represented again by the matrix $\mata{P}$
of size $N_{\rm{FFT}} \times N_{\rm{FFT}}$, then,
\begin{equation} \label{eq:muscasim_pupil}
\begin{split}
\mata{P} &= \bigcirc\circ\tilde{\mata{P}} \\
\end{split}\end{equation}
where $\tilde{\mata{P}}$ and $\bigcirc$ are defined earlier in Eq.~\eqref{eq:pavosim_pupil}
and Eq.~\eqref{eq:pavosim_circbox} respectively.

\subsection{Simulating the photon counts in a scan}

The temporal fringes in \musca\ are generated using a pair of pupils, $\mata{P}^{(1)}$
and $\mata{P}^{(2)}$. In the physical
instrument the pupils are aligned and combined coaxially at a beamsplitter to produce
two sets of output pupils. In the simulator the output pupils are modeled as two
$N_{\sigma} \times N_{\rm{STEP}}$ matrices,
\begin{equation} \label{eq:muscasim_2fs} \begin{split}
\mata{F}^{(L)} = 1 + 2a_1a_2\Re\left\{
  ((\mata{w}\cdot\mata{v})^{\mata{T}}\mata{J_{1,N_{\rm{STEP}}}}) \circ
  \mata{Q} \circ \exp\left(i\mata{\Phi}\right)
  \right\} \\
\mata{F}^{(R)} = 1 - 2a_1a_2\Re\left\{
  ((\mata{w}\cdot\mata{v})^{\mata{T}}\mata{J_{1,N_{\rm{STEP}}}}) \circ
  \mata{Q} \circ \exp\left(i\mata{\Phi}\right)
  \right\}
\end{split}\end{equation}
The matrices and their superscripts $(L)$ and $(R)$ represent the intensity of the
output pupil on the left and right side of the beamsplitter. The variables $a_1$, $a_2$
and $\mata{w}$ are previously defined in Eq.~\eqref{eq:pavosim_a} and Eq.~\eqref{eq:pavosim_w}.
The vector $\mata{v}$, which is similar to the matrix $\mata{V}$ in
Eq.~\eqref{eq:pavosim_V}, is the model fringe visibility of the source for the baseline
defined by telescopes of pupil $\mata{P}^{(1)}$ and $\mata{P}^{(2)}$.
\begin{equation} \label{eq:muscasim_v}
\begin{split}
\mata{v} &= 
\begin{bmatrix}
V(\sigma_1B_{1,2}) & V(\sigma_2B_{1,2}) & \cdots & V(\sigma_{N_{\sigma}}B_{1,2})
\end{bmatrix} \\
\end{split}\end{equation}
Each element of $\mata{Q}$ is,
\begin{equation} \label{eq:muscasim_Q}
\begin{split}
Q_{j,k} &= \frac{\sum_{u=1}^{N_{\rm{FFT}}}\sum_{v=1}^{N_{\rm{FFT}}}
  P_{u,v}^{(1,j,k)} \overline{P_{u,v}^{(2,j,k)}}}
  {\text{Area of circular aperture in }\bigcirc} \\
\end{split}\end{equation}
The real part of Eq.~\eqref{eq:muscasim_Q}, $\Re\left\{Q_{j,k}\right\}$, represents the
normalized intensity at wavenumber $\sigma_j$ due to the sum of pupils $\mata{P}^{(1)}$
and $\mata{P}^{(2)}$ at $k$-th step of a scan. $\mata{\Phi}$ is the additional phase
difference between two telescope pupils at various wavelengths,
\begin{equation} \label{eq:muscasim_Phi}
\begin{split}
\mata{\Phi} &= 2\pi\left(
  \mats{\sigma}\mata{^T}\mats{\zeta} + 
  \mata{J_{N,1}}\mats{\sigma}\mata{NZ} +
  \mats{\sigma}\mata{^T}\mata{d}
  \right)
\end{split}\end{equation}
and it is used to model the difference in piston between the pupils at each step of a
scan.

The structure of Eq.~\eqref{eq:muscasim_Phi} is very similar to Eq.~\eqref{eq:pavosim_Phi}. The
first term in Eq.~\eqref{eq:muscasim_Phi}, $\mats{\zeta}$, is the astrometric OPD per scan
of the baseline $\mata{B}_{1,2}$. Each element, $\zeta_k$, is
evaluated at time $t_{\alpha,k}$,
\begin{equation} \label{eq:muscasim_Zeta}
\begin{split}
\zeta_{k} &= \left|\hat{s}\cdot\mata{B}_{1,2}\right|_{t={t_{\alpha,k}}}
\end{split}\end{equation}
where,
\begin{equation} \label{eq:A}
\begin{split}
t_{\alpha,k} = t_{\rm{START}} + t_{\rm{STEP}}\left(\alpha N_{\rm{STEP}} + k\right)
\end{split}\end{equation}
and $\alpha$ is the number of elapsed scans before the current one and $k$ is the index
of a step within a scan. The typical value of $t_{\rm{STEP}}$ for \musca\ is 0.3ms. The
second term in Eq.~\eqref{eq:muscasim_Phi}, $\mata{NZ}$, is the optical path of the delay
line used to compensate the astrometric OPD. The matrix, $\mata{N}$, previously defined
in Eq.~\eqref{eq:pavosim_N}, denotes the refractive indices of each optical medium in the
path while the matrix, $\mata{Z}$, denotes the path length of each medium at every step
in a scan.
\begin{equation} \label{eq:muscasim_Z}
\begin{split}
\mata{Z} &= 
\begin{bmatrix}
z_1(t_{\alpha,1}) & z_1(t_{\alpha,2}) & \cdots & z_1(t_{\alpha,N_{\rm{STEP}}}) \\
z_2(t_{\alpha,1}) & z_2(t_{\alpha,2}) & \cdots & z_2(t_{\alpha,N_{\rm{STEP}}}) \\
\vdots & \vdots & \ddots & \vdots \\
z_{N_{\rm{MED}}}(t_{\alpha,1}) & z_{N_{\rm{MED}}}(t_{\alpha,2}) & \cdots & z_{N_{\rm{MED}}}(t_{\alpha,N_{\rm{STEP}}}) \\
\end{bmatrix} + 
\begin{bmatrix}
\mats{\ell} \\
\mata{0_{1,N_{\rm{STEP}}}} \\
\vdots \\
\mata{0_{1,N_{\rm{STEP}}}} \\
\end{bmatrix}
\end{split}\end{equation}
The values of elements in the first term of the RHS of
Eq.~\eqref{eq:muscasim_Z} are calculated using the method described by \citet{Tango:1990}.
The vector $\mats{\ell}$ in
Eq.~\eqref{eq:muscasim_Z} describes the relative change in the optical path length of air at
every step in a scan due to the motion of the scanning mirror.
\begin{equation} \label{eq:muscasim_l}
\begin{split}
\mats{\ell} = &\frac{L_{\rm{SCAN}}}{N_{\rm{STEP}}-1} \times \\
&\begin{bmatrix}
-N_{\rm{STEP}}/2 & -N_{\rm{STEP}}/2+1 & \cdots & -1 & 0 & 1 & \cdots & N_{\rm{STEP}}/2-1
\end{bmatrix} \\
\end{split}\end{equation}
Elements in the same column of $\mata{Z}$ have the same timestamp as an element with the
same index in $\mats{\zeta}$. The last term in Eq.~\eqref{eq:muscasim_Phi}, $\mata{d}$, 
is the user-specified offset at each step to simulate a non-optimally compensated
astrometric OPD.
\begin{equation}\begin{split}
\mathbf{d} &= 
\begin{bmatrix}
d_1 & d_2 & \cdots & d_{N_{\rm{STEP}}}
\end{bmatrix}
\end{split}\end{equation}

After the pupils are combined and have formed fringes, it is assumed that the entire
image of the pupil falls within the active area of the photodiodes and all photons are
detected. The photodetectors in \musca\ have only one pixel and are unable to resolve
any spatial variation in intensity at the image plane. Therefore only an average
intensity is recorded, hence the term $\mata{Q}$ in Eq.~\eqref{eq:muscasim_2fs}. In
addition to that the photodetectors in \musca\ are unable to resolve intensity variation
across wavelengths. Therefore the number of photon counts recorded by the
photodetectors is the sum of photons across the entire \musca\ operating bandwidth,
\begin{equation} 
\begin{split}
\mats{\mathcal{F}} &= N_{\rm{PHOTONS}} \times \mata{J_{1,N_{\sigma}}}\mata{F} + \mats{\epsilon} \\
\end{split}\end{equation}
Similar to Eq.~\eqref{eq:pavosim_Fi}, $N_{\rm{PHOTONS}}$ is a scaling factor that converts
the intensity of the output pupil to the number of photons expected from the source
given its brightness in magnitude scale and $\mats{\epsilon}$ is a noise term included
to simulate photon noise and the dark count noise of the detector.

%% file: intensity2nphotons.tex
\section{Converting magnitude scale to photon rate} \label{sec:sim_nphotons}

The scaling factor $N_{\rm{PHOTONS}}$ in Section~\ref{sec:pavosim} and Section~\ref{sec:muscasim} is
estimated based on the expected throughput of the beam combiner to be simulated, the
efficiency (Q.E.) of the APDs and
a calibrated magnitude-to-flux scaling factor, $F_\nu$, by \citet{Bessell:1979}.
The first two factors collectively describe the efficiency of the instrument, $\eta$,
which is found to be $\sim$3\% for \pavo\ and $\sim$1\% for \musca. The lower efficiency
in \musca\ is possibly due to the aluminium coated mirrors used in \susi\ and a silvered
beamsplitter in \musca. The values of $F_\nu$ at different photometric bands are listed
in Table \ref{tab:calflux}, which is reproduced from \citet{Bessell:1979}. From $F_\nu$,
the number of photons from a $m_V$ magnitude star collected by a telescope with an area
of $A_{\rm{tel}}$ m$^2$ in $\Delta t$ seconds can be estimated. Putting the factors
together,
\begin{equation} \label{eq:m2p_nphotons}
\begin{split}
N_{\rm{PHOTONS}} \simeq \;
    &\eta \times 1.51\times10^{10} \times 10^{-m_V/2.5} \times
    F_\nu \frac{\Delta(1/\sigma)}{\Delta\lambda} A_{tel} \Delta t
\end{split}\end{equation}
where the $\Delta(1/\sigma)$ is the bandwidth of the beam combiner and $\Delta\lambda$
is the bandwidth referred from Table~\ref{tab:calflux} which has an effective wavelength
close to that of the beam combiner. As an example, in the case of the \musca\ beam
combiner, $\Delta(1/\sigma) \simeq 0.29\mu$m, $\Delta\lambda = 0.15\mu$m,
$\lambda_{\rm{eff}}=0.79$, $A_{\rm{tel}}=7\times10^{-3}$m$^2$ and $\Delta t \simeq 80$ms
(256$\times$0.3ms).


%


\begin{table}
\center
\caption{Absolute flux calibration of $\alpha$ Lyrae \citep{Bessell:1979}}
\label{tab:calflux}
\begin{tabular}{c c c c}
\hline
Filter & $\lambda_{\rm{eff}}$ & $\Delta\lambda$ & Flux density, $F_\nu$ \\
band & ($\mu$m) & ($\mu$m) & ($\times10^{-23}$ W m$^{-2}$ Hz$^{-1}$) \\
\hline
U & 0.36 & 0.076$^*$ & 1.81 \\
B & 0.44 & 0.094$^*$ & 4.26 \\
V & 0.55 & 0.088$^*$ & 3.64 \\
R$_\text{C}$ & 0.64 & 0.57-0.72 & 3.08 \\
I$_\text{C}$ & 0.79 & 0.725-0.875 & 2.55 \\
\hline
\multicolumn{4}{l}{
\footnotesize{$^*$\url{http://www.astro.umd.edu/~ssm/ASTR620/mags.html}}}
\end{tabular}
\end{table}

%% file: effatm.tex
\section{Simulating the effect of turbulent atmosphere} \label{sec:sim_effatm}

A turbulent atmosphere introduces amplitude and phase
fluctuation to an otherwise plane wavefront of light from a distant star.
Although amplitude fluctuation can be simulated,
both \pavo\ and \musca\ simulators make the assumption that the atmospheric phase
fluctuation only gives rise to perturbation in the phase of a wavefront. This
approximation by discarding the amplitude (scintillation) term, also known as the
near-field approximation, holds very well for pupils larger than 2.5cm under typical
turbulence condition \citep{Roddier:1981}.

\begin{figure}
\centering
\subfloat[]{\includegraphics[width=0.4\textwidth]{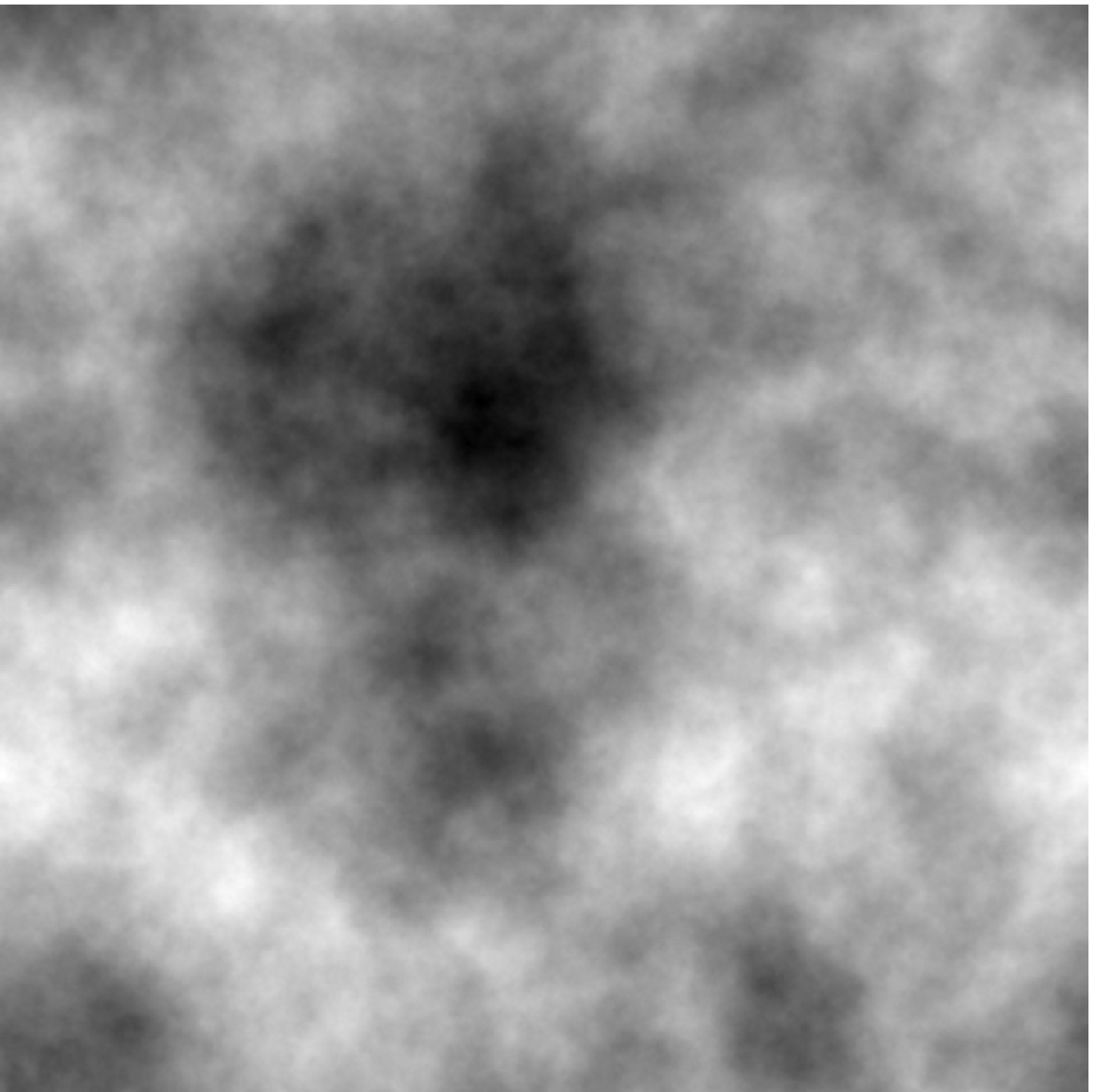}}
\hspace{1em}
\subfloat[]{\includegraphics[width=0.4\textwidth]{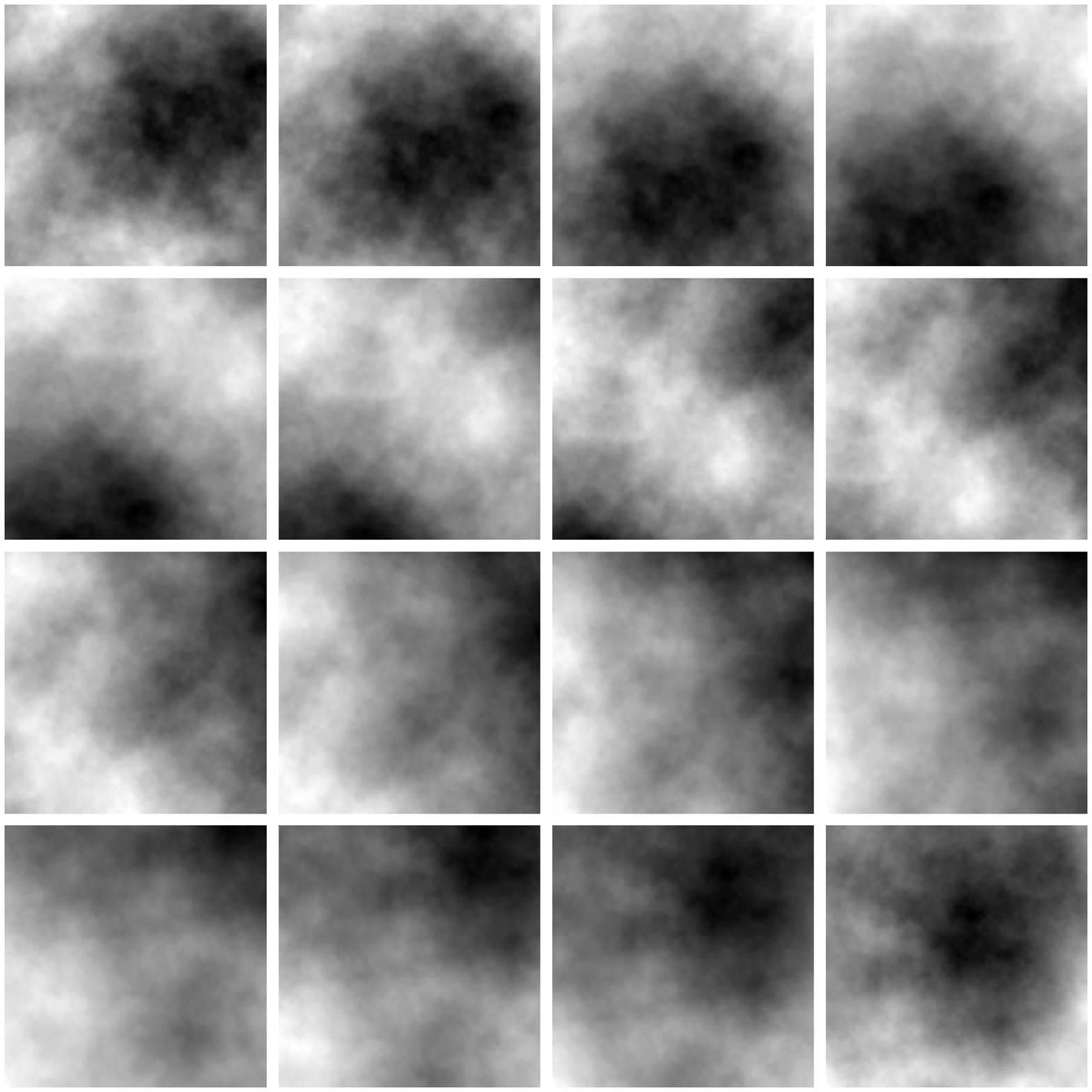}}
\caption{(a) An array of random phases generated with the inverse Fourier transform
method. The value of the phases is indicated by the shades of gray in the images. (b)
Small portions of the larger array in (a) sampled at slightly different positions at
each time step showing the progress through time. Time increases from left to right and
top to bottom.}
\label{fig:atm_screens}
\end{figure}

The phase fluctuation in the atmosphere is simulated using a large ($N_{\rm{ATM}} \times
N_{\rm{ATM}}$) two-dimensional array of random phasor, $\mats{\phi}$, which has a power
spectrum given by \citep{Roddier:1981, Glindemann:2011}, 
\begin{equation} \label{eq:atm_hatphi_el}
\begin{split}
\left|\hat{\phi}_{u,v}\right|^2 &= C \left(L_0^{-2}+(u^2+v^2)\right)^{-11/6}
\end{split}\end{equation}
The notation $\fts{\phi}$ denotes the Fourier transform of $\phi$, $u$ and $v$ are the
indices of the array and $L_0$ is the outer scale of the turbulent structure of the
model atmosphere, which is set to a very large number in the simulation. The phase
fluctuation is recovered by taking the inverse Fourier transform of
$\left|\hat{\phi}_{u,v}\right|\exp(i\varepsilon)$ where $\varepsilon$ is
the phase of the Fourier transform and it is a random variable. The randomness of the
generated phase fluctuation, $\phi$, is controlled by adjusting the scaling factor, $C$.
It is tweaked so that the structure function of $\phi$,
\begin{equation} \label{eq:atm_ddef}
\langle \left|\phi_{u,v}-\phi_{u+r_u,v+r_v}\right|^2 \rangle = 
  6.88\left(\frac{\sqrt{r_u^2+r_v^2}}{r_0}\right)^{5/3}
\end{equation}
has its characteristic Fried parameter, $r_0$, set to the value specified by the user.
Fig.~\ref{fig:atm_screens}(a) shows an image of the random phases generated using this
method. The grayscale of the images indicate the value of the phases.

On the other hand the phase fluctuation of a wavefront across a telescope, or
$\mats{\varphi}$ in Eq.~\eqref{eq:pavosim_pupil} and Eq.~\eqref{eq:muscasim_pupil}, is
simulated by extracting a small portion ($\sim$100$\times$100) of the much larger
$\mats{\phi}$ array. The typical size of the $\mats{\phi}$ array is 2048$\times$2048.
Since $r_0$ is specified at a certain wavelength, $\sigma_{r_0}^{-1}$,
and the phase fluctuation in $\mats{\phi}$ is
generated according to $r_0$, the value extracted from the larger array must be scaled
by a factor of $\sigma_j/\sigma_{r_0}$, where $\sigma_j$ is the desired wavenumber for
simulation, before applying it to the model pupils in Eq.~\eqref{eq:pavosim_pupil} and
Eq.~\eqref{eq:muscasim_pupil}. The position of this small sampling window is displaced
across the larger array after every time step of the simulation. This simulates the
effect of \citeauthor{Taylor:1938}'s \citeyearpar{Taylor:1938} hypothesis of a frozen
atmosphere drifting across the aperture of the telescope. The rate of displacement of
the sampling window depends on the wind speed which is estimated from,
\begin{equation} \label{eq:wind}
\bar{v} = 0.314\frac{r_0}{\tau_0}
\end{equation}
where $\tau_0$ is the coherence time of the phase fluctuation and the time step of the
simulation, $t_{\rm{STEP}}$. Both parameters are specified by the user. The direction of
the displacement is random but remains constant throughout the simulation.
Fig.~\ref{fig:atm_screens}(b) shows several snapshots of phase variation over a
small portion of a larger array drifting across the sampling window.

A separate array of $\mats{\phi}$ is generated for each telescope. By doing this, it is
assumed that the phase fluctuations over individual telescopes are uncorrelated. As an
effect, the low-frequency phase fluctuations do not increase with baseline in the
simulation.
In practice this scenario is true if the baseline is longer than the outer scale of
turbulence, $L_0$, which is in the order of 100m in the troposphere \citep{Roddier:1981}.
Therefore this approach of having a separate array of $\mats{\phi}$ for each telescope
does not simulate the piston term of an aberrated pupil due to atmospheric turbulence
realistically. In order to address this shortcoming in the \pavo\ and \musca\ simulator,
the input variable, $\mata{D}$ or $\mata{d}$ (refer Table~\ref{tab:pavosim_inputs}), can
be used to add an extra differential piston between pupils from two apertures.

\begin{figure}
\centering
\includegraphics[width=\textwidth]{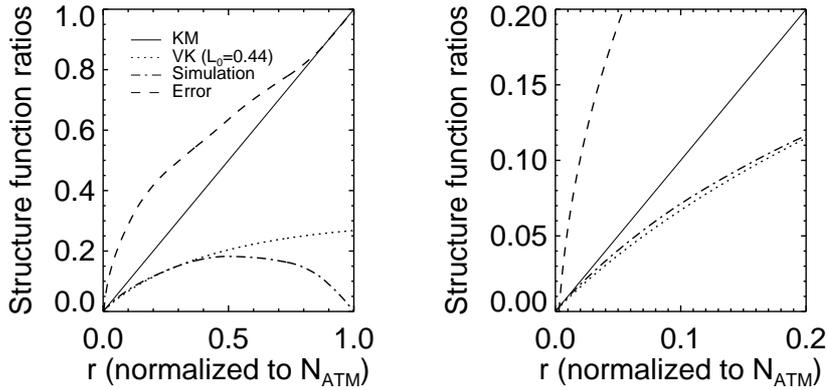}
\caption{Plots of $(D_\phi/6.88)^{3/5}$ against $r$ where $D_\phi$ is either a
theoretical, based on a Kolmogorov (KM) model, Eq.~\eqref{eq:atm_ddef}, or a von
Karman (VK) model \citep{Valley:1979}), and a simulated structure function. Both axes
are normalized to the size of the random phases array. The dashed lines in the plots are
the difference between the solid (KM) and dash dot (simulation) lines
relative to the former. If the simulated structure function is exactly the same as the
theoretical structure function then the dash lines will be horizontal at 0. The right
plot is the zoomed in version of the left.}
\label{fig:atm_check}
\end{figure}

Another shortcoming of this method in simulating the phase variation across a
turbulent atmosphere is a structure function that deviates from its theoretical value at
large distances. Fig.~\ref{fig:atm_check} shows comparison between a theoretical
structure function and one calculated from the $\mats{\phi}$. The plots in the figure
are $(D_\phi/6.88)^{3/5}$ versus $r$ and normalized to the size of the array,
$N_{\rm{ATM}}$. The theoretical Kolmogorov (KM) model curve is a straight line with
unity slope ($r_0=1$). The deviation from the theoretical value is very pronounced
especially at large distances. This is not surprising because the power spectrum,
$|\fts{\phi}|^2$, is undersampled at low frequencies, where most of the energy resides.
A practical solution to this problem, which is the approach implemented in this work, is
to sub-sample the phase fluctuation from a much larger array
\citep{McGlamery:1976,Shaklan:1989,Lane:1992}. With a telescope aperture model of
$\sim$5\% of the size of the larger phase array, such approximation produces phase
fluctuation that is at most 20\% off the theoretical KM model. This is shown in
Fig.~\ref{fig:atm_check}. Also shown in the figure is a theoretical von Karman (VK)
model which has a structure function very similar to that of the generated phase
fluctuation. The outer scale of turbulence of the model is $\sim$44\% of
$N_{\rm{ATM}}$ and at this outer scale of turbulence, given the telescope aperture used
in the simulation model, the variance of the tip-tilt angle of an image formed with the
phase fluctuation is reduced by $\sim$40\% from its expected value based on a KM model
which has an infinite outer scale \citep{Sasiela:1993}. This reduction in tip-tilt
fluctuation simulates the active tip-tilt correction at \susi\ which stabilizes the
image of a star.

%% file: testcases.tex
\section{Testcases} \label{sec:sim_testcases}

The main objective of this simulation framework is to test the data reduction pipeline
of the beam combiners. It is also a good tool to investigate software bugs in a data
reduction pipeline during its development stage. The following sections discuss some
testcases using the simulators.

Several testcases were carried out to demonstrate the functionality of the
simulator and at the same time the accuracy of the data reduction pipeline for both
\pavo\ and \musca.
Testcase~I is carried out to verify the extraction of visibility squared, $V^2$, of
a set of fringes by the \pavo\ pipeline.
Testcase~II-IV were carried out to probe the lower bound phase error of the
phase-referenced fringes constructed by the \pavo\ and \musca\ phase-referencing
pipeline. On top of that they are also used to verify the \pavo\ $V^2$ pipeline.
Table~\ref{tab:tc_bothsim_inputs} shows the input for each testcase. All input
parameters for Testcase II-IV are kept the same except for the fringe visibility
parameter.

\input{testcases_results.tex}

\subsection{To verify the \pavo\ $V^2$ reduction pipeline} \label{sec:testcase1}
\begin{figure}
\centering
\includegraphics[width=0.9\textwidth]{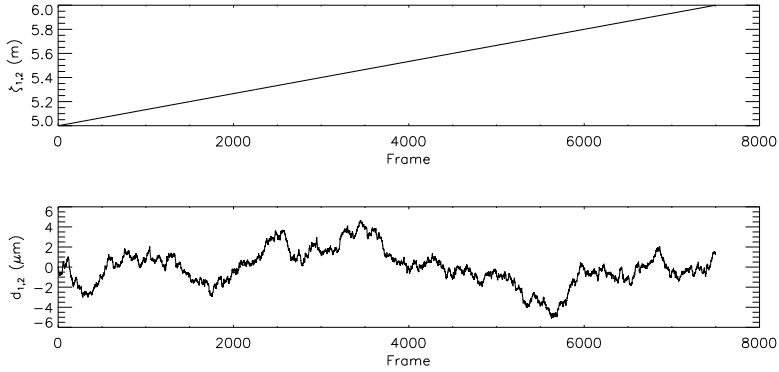}
\caption{The value of $\zeta_{1,2}$ and $d_{1,2}$ used in the simulation versus the
frame number. The latter is a pseudo-random variable where the difference in value
between each step is $d_{1,2}(t_\alpha)-d_{1,2}(t_{\alpha-1}) \sim \mathcal{N}(0,0.01) \mu$m.}
\label{fig:tc_pavosim_zeta_D}
\end{figure}

Testcase~I-IV are used to verify the $V^2$ reduction pipeline of \pavo\ at \susi. In
these testcases the \pavo\ simulator is used to investigate the pipeline's ability to
reproduce the square of the visibility of a given model.
The input fringe visibility for Testcase I is a sinusoidal function which models
a binary star system with the primary and secondary stars almost equal in
brightness (contrast ratio of $\sim$0.95) and have a projected separation of
$\sim$0.04$''$ using a 15m baseline.
On the other hand, the input fringe visibilities for Testcase II-IV are constant
to model an unresolved single star but have different values of instrument
visibilities.
Since \pavo\ at \susi\ uses
only two telescopes at any one time, $\mats{\zeta}$ and $\mata{D}$
each reduces to a $2\times2$ square matrix. The values of off-diagonal elements of the
matrices, $\zeta_{1,2}$ and $d_{1,2}$, are shown in Fig.~\ref{fig:tc_pavosim_zeta_D}.
All testcases have 3 sub-cases (A, B and C) where the photon rate is varied to simulate
an observation of a zeroth, 2nd or 4th magnitude star. The number of photons in a
generated \pavo\ frame is adjusted to match the number in an actual frame recorded by
the \pavo\ camera at a gain of 5, for a star of magnitude $m_V=0.0$ and $m_V=2.0$, and
25, for a star of magnitude $m_V=4.0$.

\begin{figure}
\centering
\subfloat[]{\includegraphics[width=0.5\textwidth]{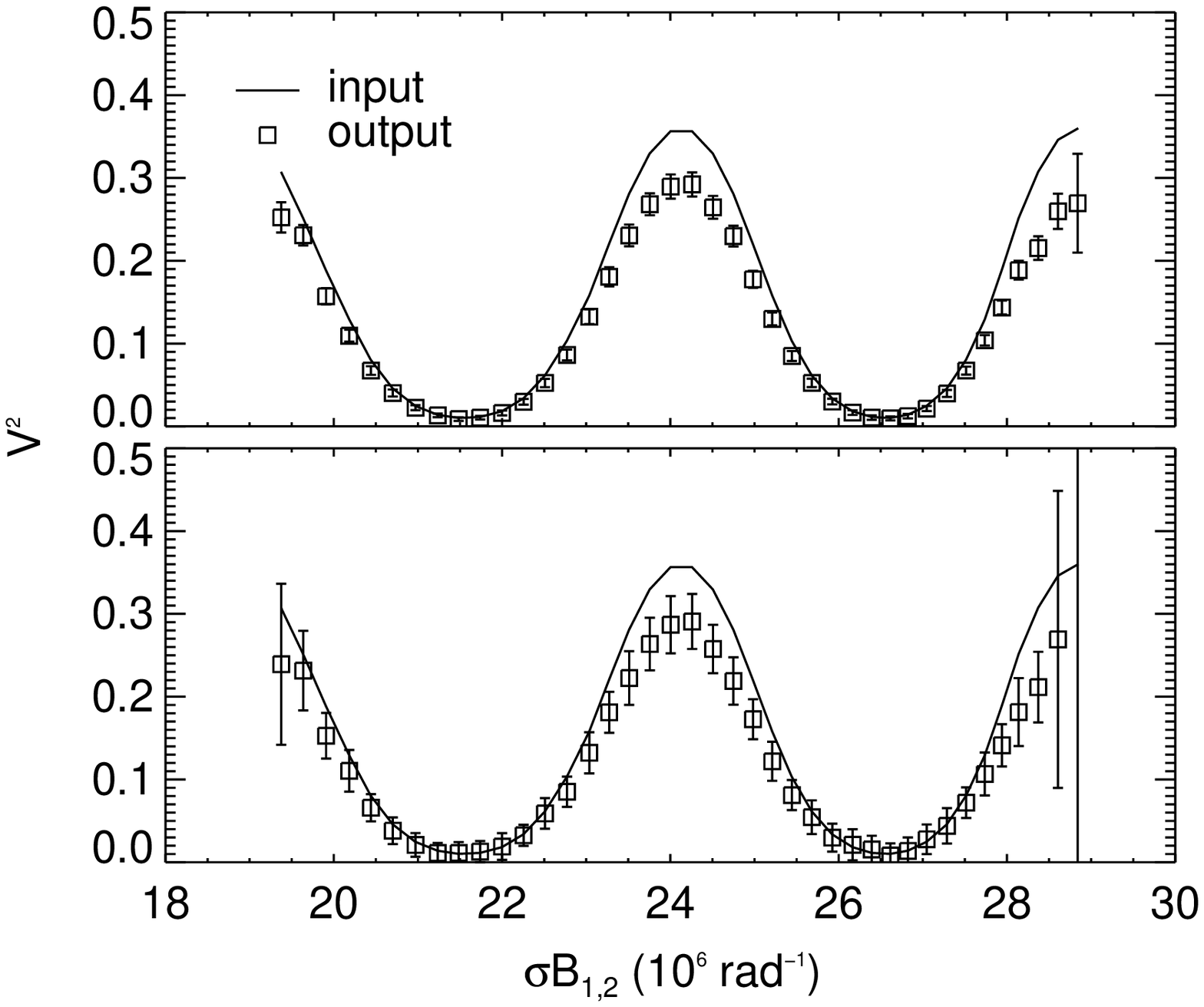}}
\subfloat[]{\includegraphics[width=0.5\textwidth]{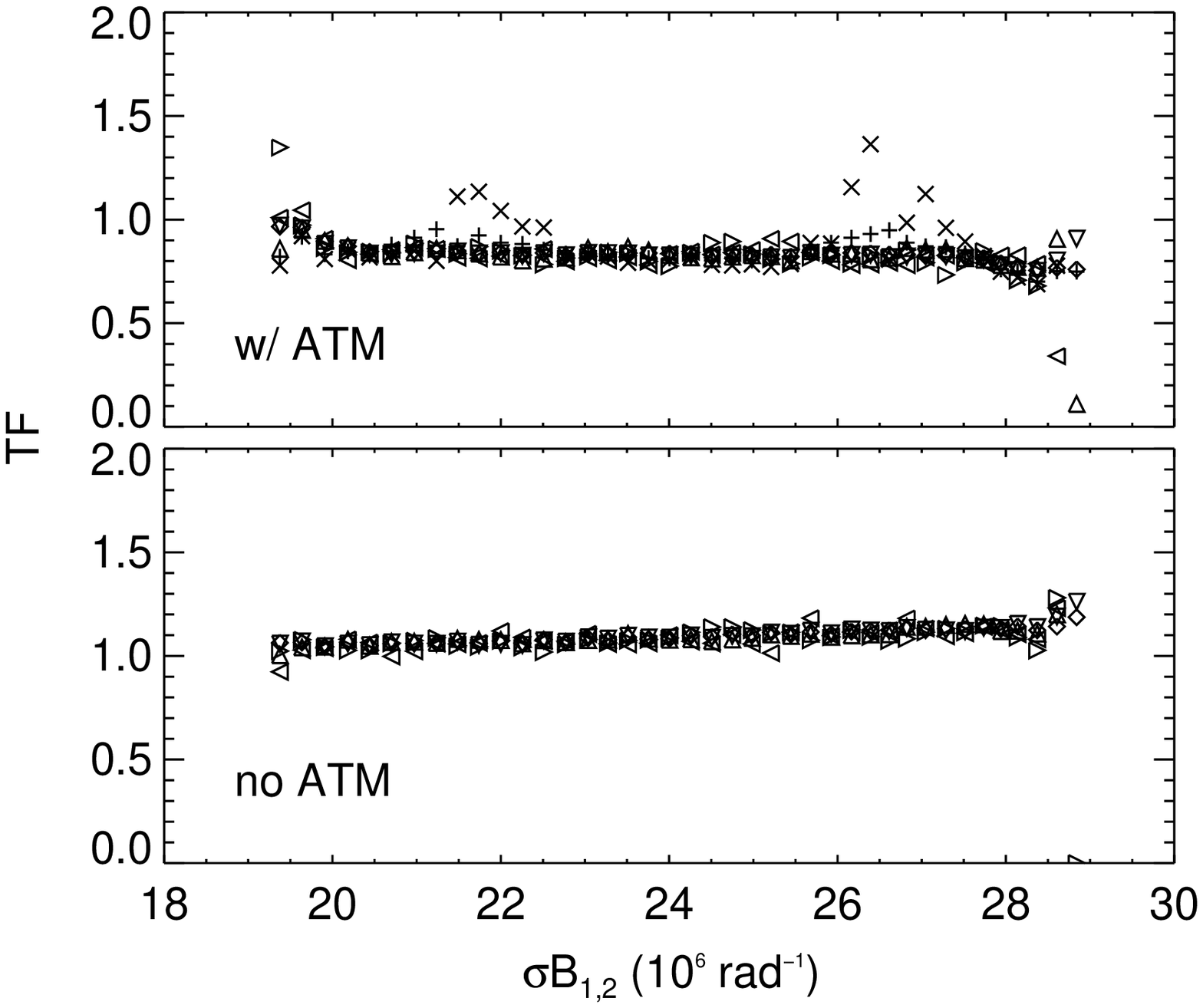}} \\
\caption{A comparison between user specified models of $V^2$ (solid line) and the
estimated $V^2$ produced by the \pavo\ $V^2$ reduction pipeline. The top plot of (a)
shows the comparison between a wavelength dependent $V^2$ model (input) and the (output)
values estimated by the pipeline at high photon rate (sub-case A) while the bottom plot
of (a) represents a similar comparison at lower photon rate (sub-case B). The ratio of
estimated to model $V^2$ for Testcase~I~($+,\times$), II~($\diamond$,$\vartriangle$),
III~($\triangledown$,$\triangleright$) and IVB~($\triangleleft$) are plotted against
spatial frequency in (b). Each pair of symbols in the parentheses except for
Testcase~IV represent sub-case A and B respectively. Also in (b), the ratios are
shown when atmospheric phase noise was excluded in the simulation (no ATM).}
\label{fig:tc1}
\end{figure}

\begin{figure}
\centering
\subfloat[]{\includegraphics[width=0.9\textwidth]{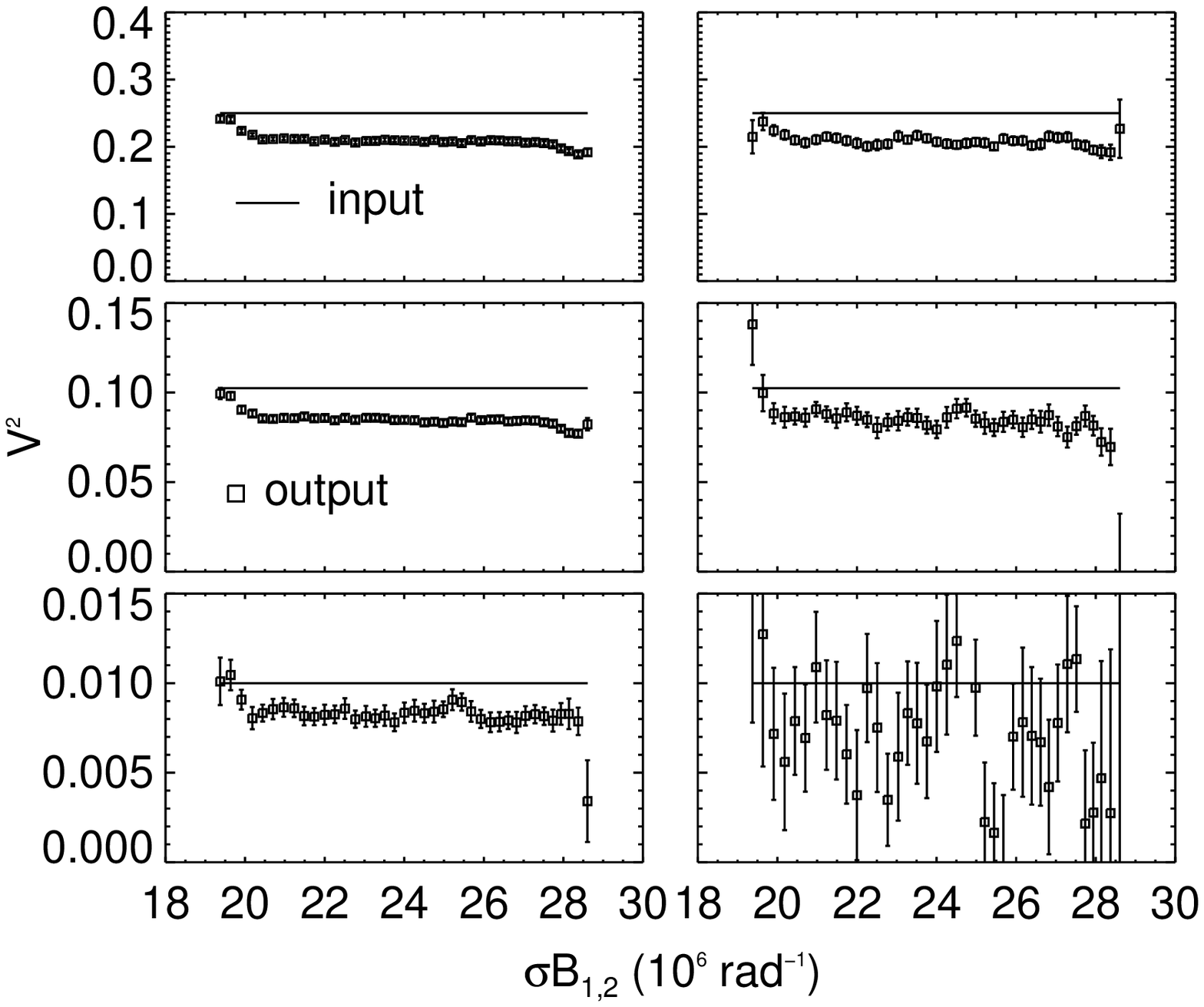}} \\
\vspace{-2em}
\subfloat[]{\includegraphics[width=0.9\textwidth]{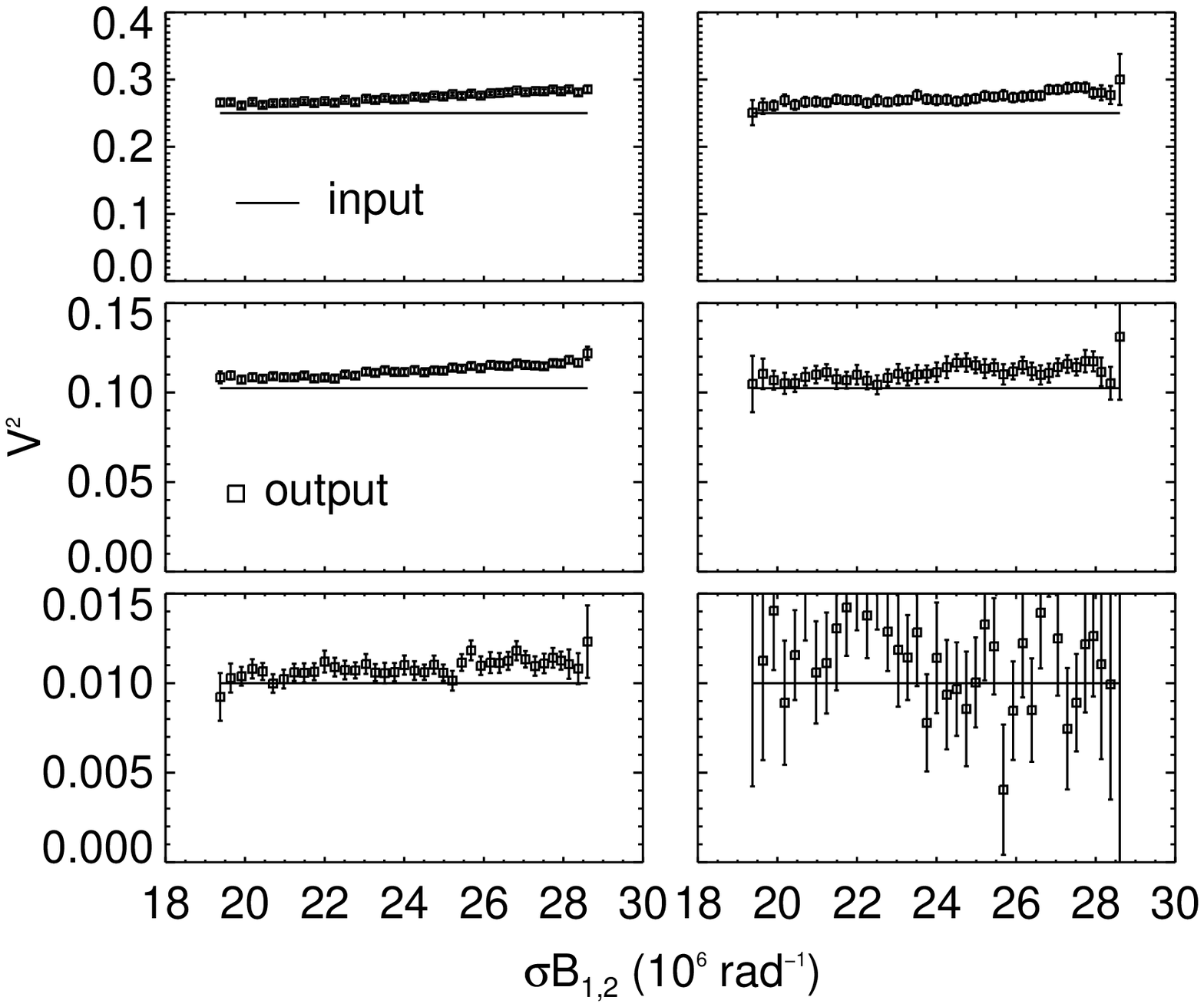}}
\caption{Similar to Fig.~\ref{fig:tc1}(a), the top, middle and bottom plots of (a) and
(b) show the comparison between a constant $V^2$ model (input) and the estimated
(output) of the pipeline for Testcase~II, III and IV respectively. The plots on the left
column are from sub-case A where the photon rate is higher than sub-case B which plots
are on the right column. The estimated $V^2$ values of sub-case B of Testcase IV are
very noise but are still unbiased. The plots in (a) are simulated with the
presence of atmospheric phase noise while plots in (b) are simulated without.}
\label{fig:tc234}
\end{figure}

The four types of frames generated by the \pavo\ simulator are similar to those shown in
Fig.~\ref{fig:pavosim_fits}(c) and therefore will not be shown again in this section. The
estimated $V^2$ of each simulated data set extracted by the \pavo\ pipeline are plotted
against the models fed into the simulator in Fig.~\ref{fig:tc1}. Also plotted for
comparison, in Fig.~\ref{fig:tc234}, are the estimated $V^2$ of Testcase II-IV.
Both figures show that the estimated values are consistent and within 20\% of the
model. The expected reduction in visibility of the fringes, which is more
pronounced at higher spatial frequencies (right half of the graphs), is due to
the atmospheric phase noise.
The observed trend in Fig.~\ref{fig:tc234}(b), is due to a wavelength-dependent
scaling factor in the order of unity which is in turn
caused by the shape of a Fourier domain windowing function  used in the data
reduction pipeline. This scaling factor can be calibrated out because it is
identical regardless of the visibility function, as seen in
Fig.~\ref{fig:tc1}(b) which plots the ratios between the estimated and the model
$V^2$ (or TF) across spatial frequency.
Data points with TF
significantly larger than 1 is an effect of calculating a ratio with the
denominator that has a very small value.
%

In summary, these testcases show that the $V^2$ reduction pipeline has some dependence
on the seeing condition, which can be calibrated out, but no measurable bias as a
function of target brightness and the square of the visibility of the fringes.

\subsection{To verify the \pavo\ and \musca\ phase-referencing pipeline} \label{sec:testcase2}
In addition to the goal mentioned in previous section, the main aim of Testcase~II-IV is
to test a newly written \pavo\ and \musca\ phase-referencing pipeline and to probe the
lower bound phase error of the phase-referenced fringes. Both simulators are fed
with the same input for this purpose.  Some inputs which are common to both simulators,
e.g.\ $\mats{\zeta}$ and $\mata{D}$ for \pavo, are resampled at a different rate for
\musca\ because the time steps of the two simulators are different. Inputs which are not
common to both inputs, e.g.\ are defined separately.

\begin{figure}
\centering
\subfloat[]{\includegraphics[width=0.5\textwidth]{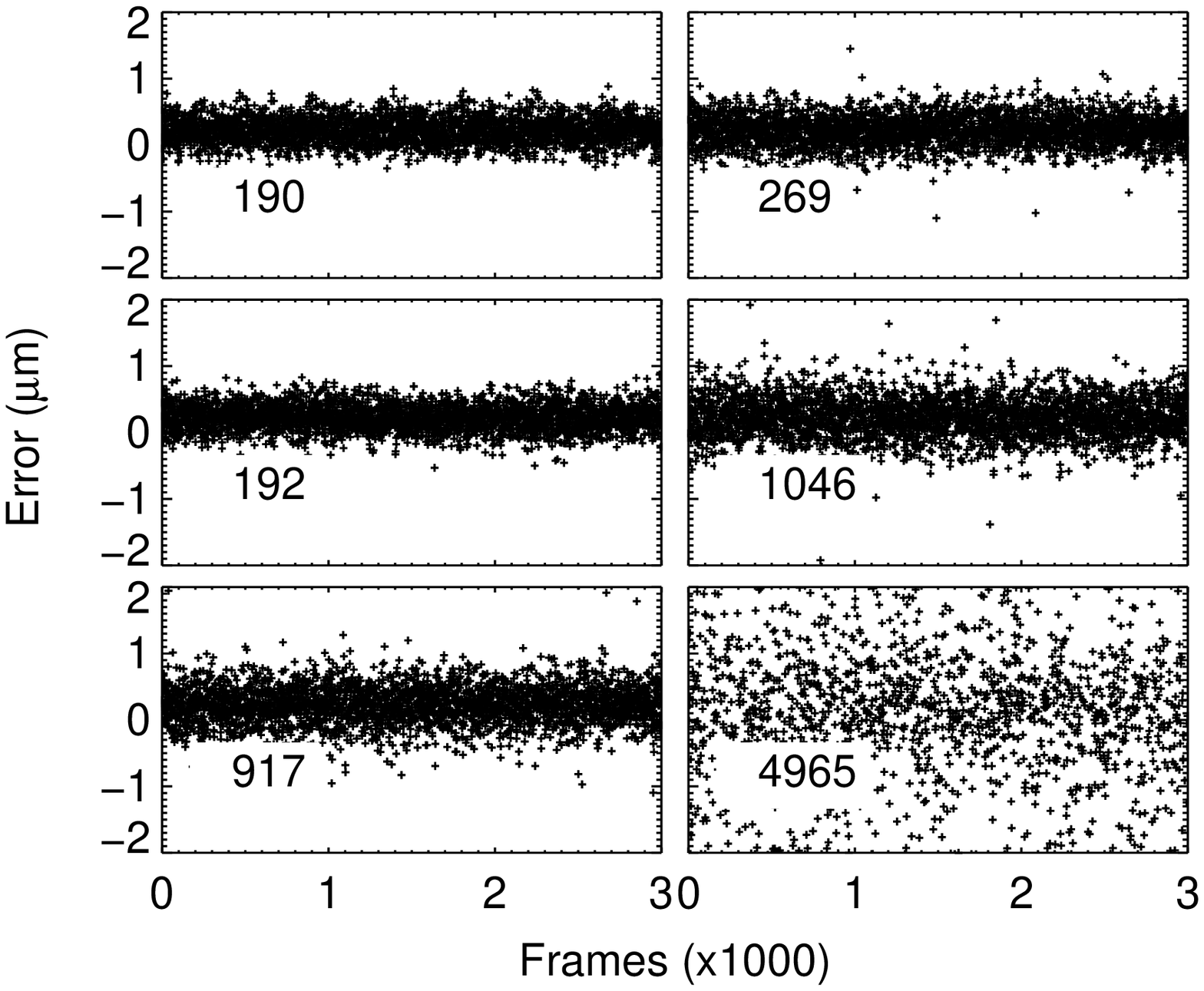}}
\subfloat[]{\includegraphics[width=0.5\textwidth]{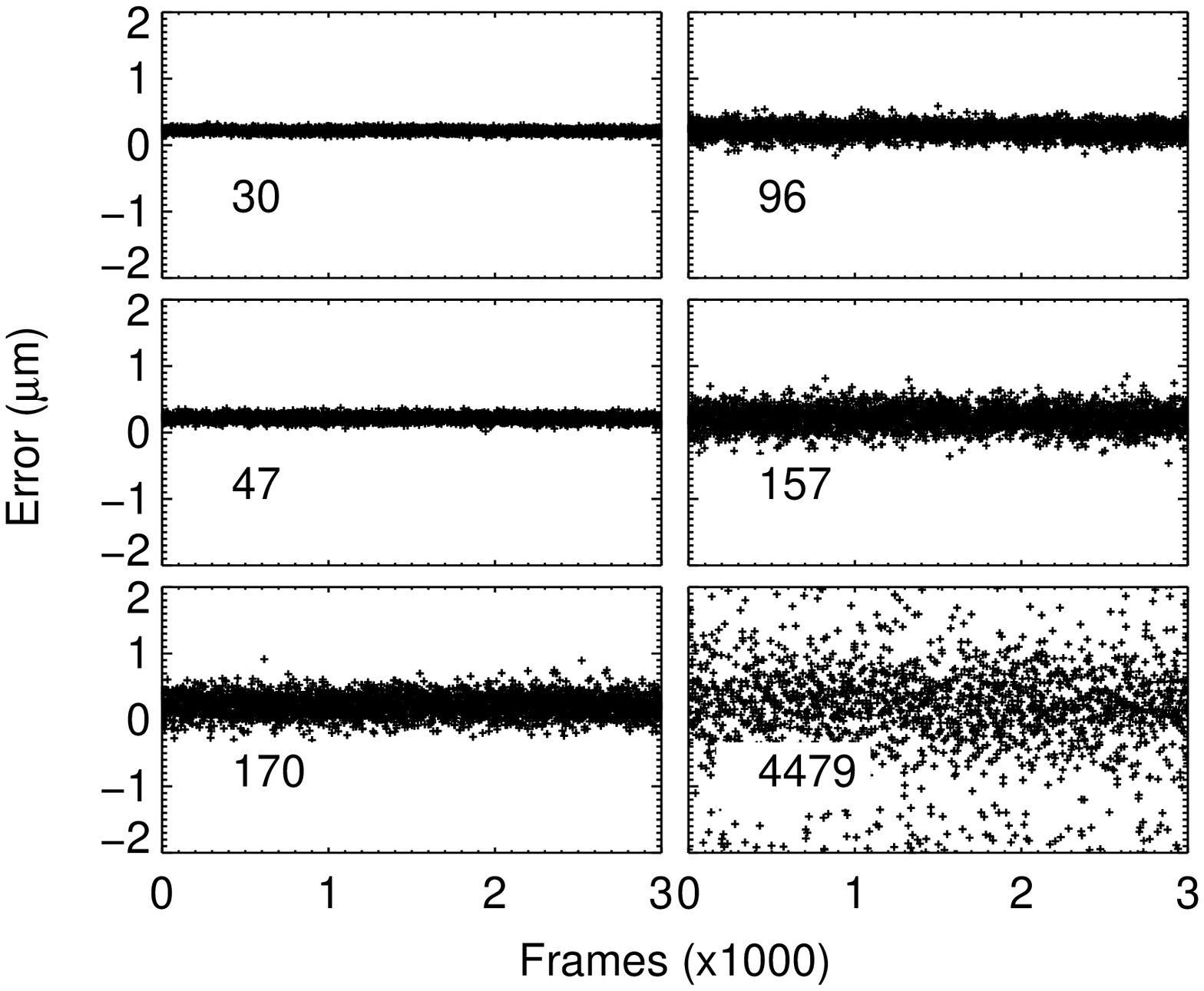}} \\
\caption{The errors of group delay estimates obtained from the \pavo\ reduction pipeline
for Testcase II-IV (from top to bottom). The left and right columns are errors for
sub-cases A and B respectively. The standard deviation of the errors, in unit of
nanometers, are stated in the plots. The data for the plots in (a) and (b) are obtained
with and without atmospheric phase noise in the simulations.}
\label{fig:tc234_opdres}
\end{figure}

The role of the \pavo\ data reduction pipeline in these testcases is to provide
estimates of the group and phase delay of the fringes. The group delay of the fringes
is defined by the user-specified input variables, $\mata{D}$ and $\mata{d}$. The
errors of the group delay estimates are shown in Fig.~\ref{fig:tc234_opdres}. The
standard deviation of the error of the group delay estimates increases as the
visibility of the fringes and the photon rate decrease. This is expected as the signal
becomes weaker than the photon noise and read noise of the camera. Since the phase delay
of the fringes is estimated from the group delay estimate, residuals of group delay
errors which are larger than one wavelength of the fringes produce unreliable phase
delay estimates. Only reliable estimates of phase delay are chosen and applied to
stabilize the position of a fringe packet generated by the \musca\ simulator
\citep{Kok:2012}. The percentage of reliable estimates within an observation is a
function atmospheric seeing, brightness of the target star and the visibility of the
stellar fringes.

\begin{figure}
\centering
\subfloat[]{\includegraphics[width=0.5\textwidth]{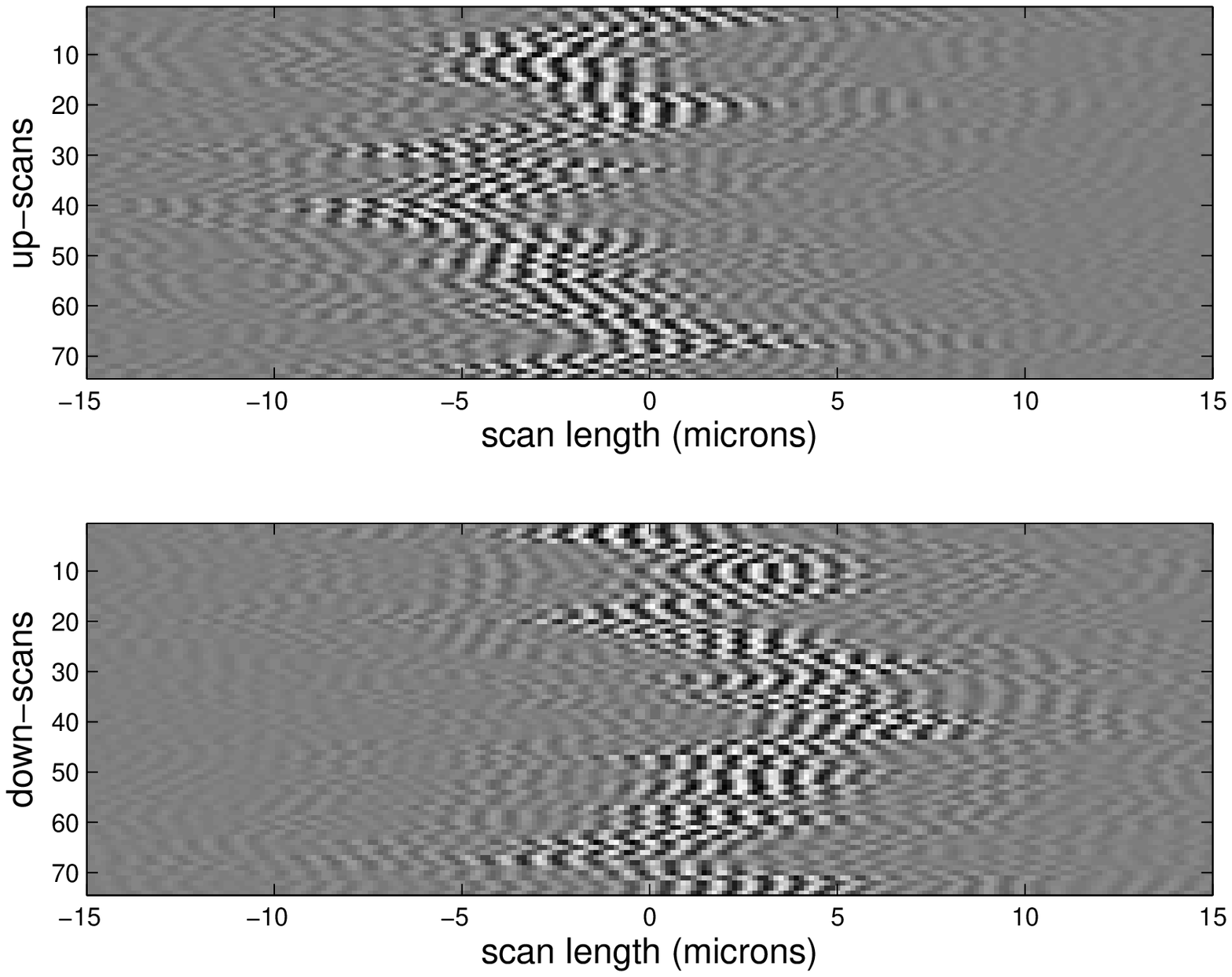}}
\subfloat[]{\includegraphics[width=0.5\textwidth]{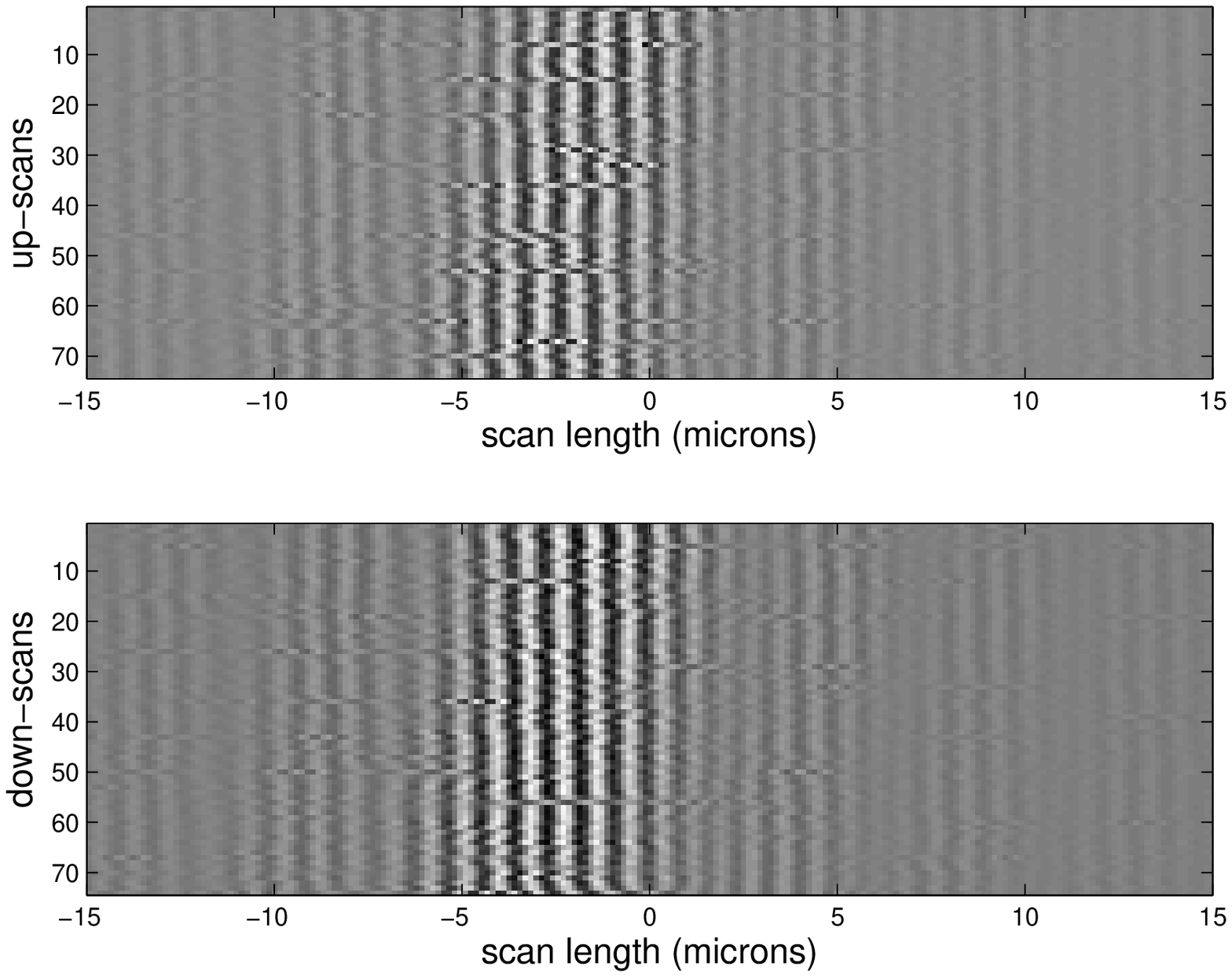}}
\caption{Simulated \musca\ fringes in a waterfall plot format showing the effect of (a)
without and (b) with phase-referencing. The vertical axis is time and the horizontal
axis is the optical delay. The top and bottom plots are of different scan direction.}
\label{fig:tc234_waterfall}
\end{figure}

Waterfall plots in Fig.~\ref{fig:tc234_waterfall} show the position of the fringe packet
through time. They clearly show that the fringe packet has a continuously changing
position and a relatively constant position before and after the phase delay estimates
are applied. The phase variation of the phase-referenced fringes across multiple scans,
as seen in the plot, can be estimated from the $\langle|C|^2\rangle$ metric, which is a
measure of the coherence of the fringes \citep{Kok:2012}, and is given as,
\begin{equation} \label{eq:tc234_sx}
\begin{split}
s_x^2 &\approx -2\ln{\langle|C^2|\rangle}\frac{N_{\rm{SC}}}{N_{\rm{SC}}-1}
  \left(\frac{1}{2\pi\tilde{\sigma}_{\rm{M}}}\right)^2
\end{split}\end{equation}
where $N_{\rm{SC}}$ is the number of good scans used in the estimation and
$\tilde{\sigma}_{\rm{M}}$ is the mean wavenumber of \musca\ fringes, which has a
typical value of 1.2$\mu$m$^{-1}$. The factor $\frac{N_{\rm{SC}}}{N_{\rm{SC}}-1}$ is
used to obtain an unbiased estimate of the sample variance. Table~\ref{tab:tc234_c22rms}
shows the value of $\langle|C|^2\rangle$, $N_{\rm{SC}}$ and $s_x$ for all testcases.

\begin{table}
\center
\caption{Results from \pavo\ and \musca\ phase-referencing algorithm}
\label{tab:tc234_c22rms}
\begin{tabular}{c c c c c c c c}
\toprule
                      & & \multicolumn{3}{c}{with ATM} & \multicolumn{3}{c}{without ATM} \\
\cmidrule{3-8}
Testcases $\downarrow$ & $m_V$ $\rightarrow$ & 0.0 & 2.0 & 4.0 & 0.0 & 2.0 & 4.0 \\
\midrule
\multirow{3}{*}{II ($V^2=0.25$)}
			& $N_{\rm{SC}}$			& 74		& 74		& 72
							& 74		& 74 		& 74\\
			& $\langle|C|^2\rangle$		& 0.6407	& 0.3740	& 0.4112 
							& 0.8236	& 0.5678	& 0.5067\\
			& $s_x$ (nm)			& 126		& 188		& 179
							& 84		& 143		& 156\\
\midrule
\multirow{3}{*}{III ($V^2=0.10$)}
			& $N_{\rm{SC}}$			& 74		& 58		& 21
							& 74		& 74		& 74\\
			& $\langle|C|^2\rangle$		& 0.6387	& 0.2630	& 0.1352
							& 0.7582	& 0.3752	& 0.3708\\
			& $s_x$ (nm)			& 127		& 220		& 275 
							& 100		& 188 		& 189\\
\midrule
\multirow{3}{*}{IV ($V^2=0.01$)}
			& $N_{\rm{SC}}$			& 58		& 0		& 0
							& 74		& 0 		& 0\\
			& $\langle|C|^2\rangle$		& 0.2684	& N/A		& N/A
							& 0.3346	& N/A 		& N/A\\
			& $s_x$ (nm)			& 218		& N/A		& N/A
							& 198		& N/A 		& N/A\\
\bottomrule
\end{tabular}
\end{table}

For a scan to be considered good, there must be a continuously reliable phase delay
estimate throughout at least $\frac{3}{4}$ of the time to make the scan. The number of
good scans as well as the value of $\langle|C|^2\rangle$ decrease in the same trend as
the standard deviation of the errors of the group delay estimates increases because
residuals of the errors which are larger than one wavelength of the fringes produce
unreliable phase delay estimates. Testcase~IVB is an example of an extreme case as there
are no good scans found.

\begin{figure}
\centering
\subfloat[]{\includegraphics[width=0.5\textwidth]{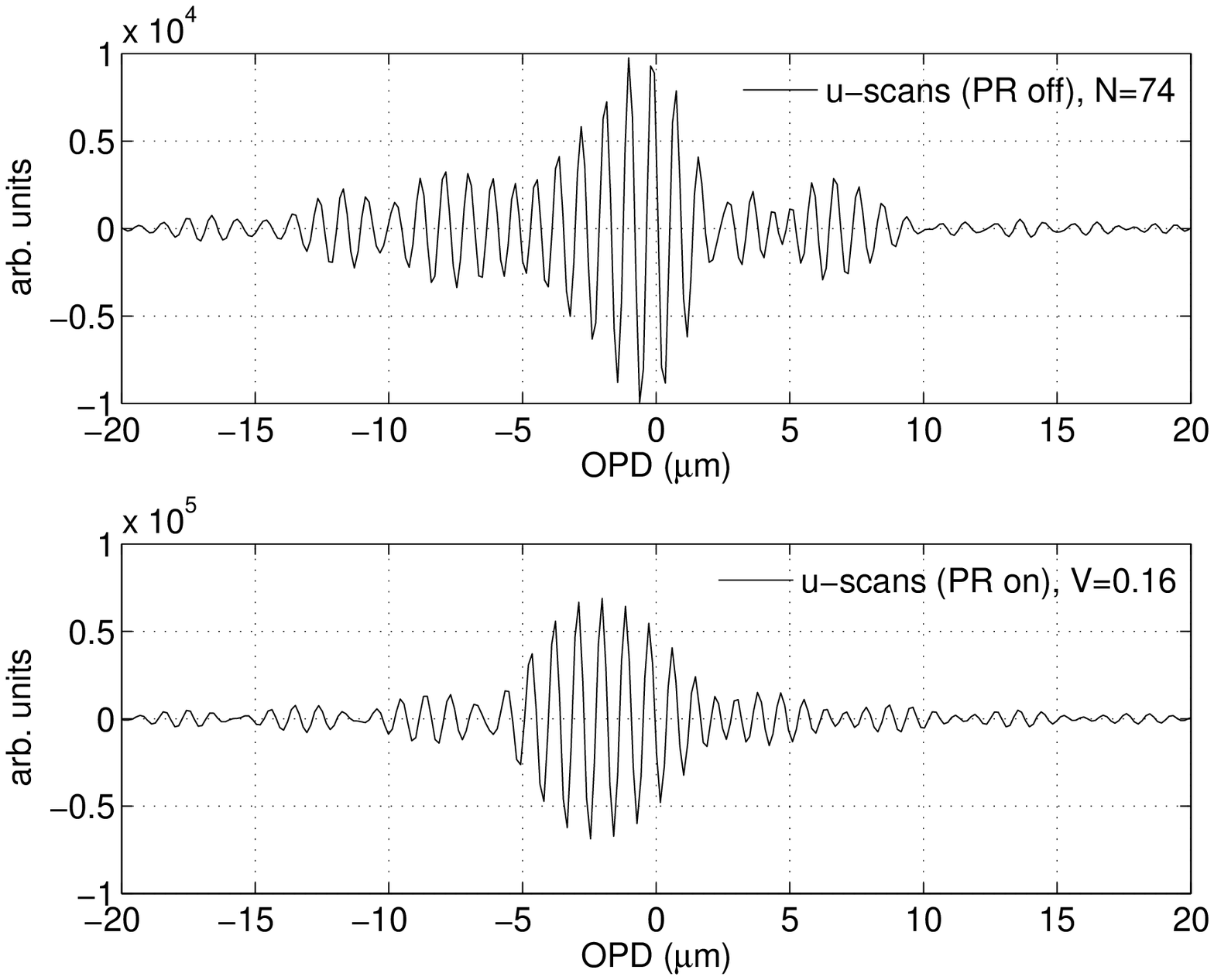}}
\subfloat[]{\includegraphics[width=0.5\textwidth]{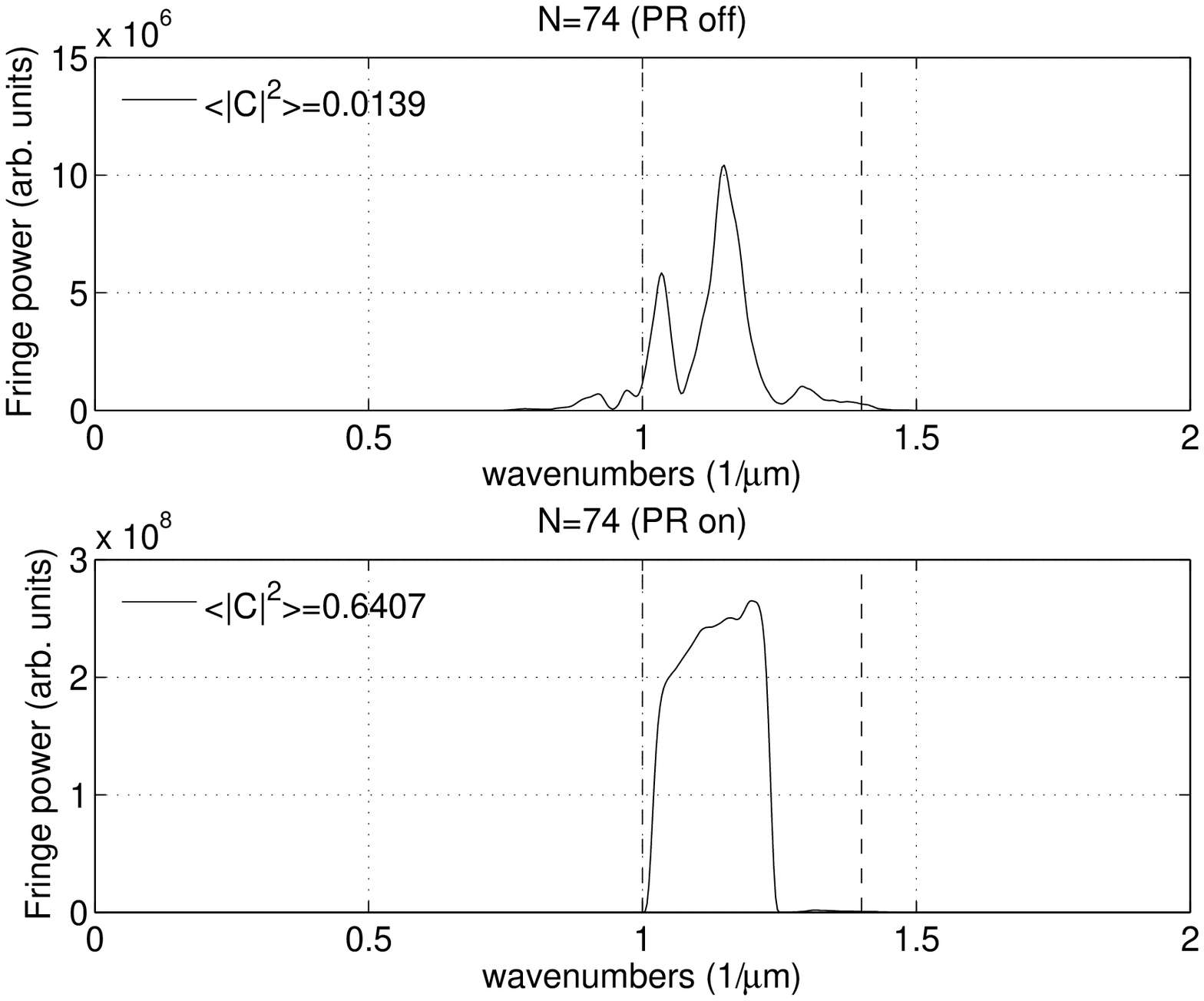}}
\caption{The top and bottom plots of (a) show the result of summation of multiple scans
of incoherent and coherent fringes respectively. Similarly the top and bottom plots of
(b) show the huge (about an order of magnitude) difference in the power spectrum of the
incoherently and coherently summed fringe packets respectively. The number of scans used
for the summation and the $\langle|C|^2\rangle$ metric are indicated in (b).}
\label{fig:tc234_befaft}
\end{figure}

The summation of the phase-referenced fringes in all good scans reproduces the fringe
packet without the distortion introduced by the atmospheric phase noise. The effect of
coherent summation of the fringes is shown in Fig.~\ref{fig:tc234_befaft}. The position
of the fringe packet within the scan range of the scanning mirror can be determined to
an uncertainty defined by $s_x/\sqrt{N_{\rm{SC}}}$. For example, in order to determine
the fringe packet position to an uncertainty of 5nm, which translates to an astrometric
uncertainty of 10$\mu$as with 100m baseline, in Testcase~IIIB, at least 1900 good scans
are required. This number is not impractical as a period of good seeing at \susi\ can
produce over 1000 good scans \citep{Kok:2012} in about 10 minutes of observation (a
total of 3600 scans) with $\beta$~Cru, which has a V magnitude of 1.3 and an
uncalibrated $V^2$ range of 0.02--0.20 as measured by \pavo.

Although the results in Table~\ref{tab:tc234_c22rms} were simulated for the setup at
\susi, they can be extrapolated to estimate the performance of a similar setup at
different observation sites. For example, the scenario in Testcase~IIC (4th
magnitude star, seeing condition of $\sim$1$''$ and diameter of light collecting
aperture of 14cm) is equivalent to observing a $\sim$6th magnitude star at a
site with $\sim$0.5$''$ seeing (e.g.\ at NPOI\footnote{Navy Precision Optical
Interferometer} or in Antarctica) but with twice the aperture size (28cm). In
such scenario, at least 1000 good scans are required to achieve an astrometric
precision of 10$\mu$as with a 100m baseline. The phase-referencing performances
in Testcase~IIB and IIC are similar because a higher EMCCD gain was used to
compensate the lower flux in the latter.

%% file: testcases_results.tex
\begin{table}
\center
\caption{Input parameters for all testcases}
\label{tab:tc_bothsim_inputs}
\begin{tabular}{c c c c c}
\toprule
Testcases $\rightarrow$ & \multirow{2}{*}{I}
			& \multirow{2}{*}{II}
			& \multirow{2}{*}{III}
			& \multirow{2}{*}{IV} \\
Names $\downarrow$      & & & & \\
\midrule
Section			& Section~\ref{sec:testcase1} &
\multicolumn{3}{c}{Section~\ref{sec:testcase1} \& \ref{sec:testcase2}} \\
$t_{\rm{START}}$	& \multicolumn{4}{c}{2455848.5} \\
\cmidrule{2-5}
\multirow{2}{*}{$t_{\rm{STEP}}$}
			& \multicolumn{4}{c}{\pavo: 5.0ms} \\
                	& \multicolumn{4}{c}{\musca: 0.2ms} \\
\cmidrule{2-5}
$N_{\rm{S-FITS}}$	& \multicolumn{4}{c}{150} \\
$N_{\rm{R-FITS}}$	& \multicolumn{4}{c}{20}  \\
$N_{\rm{F-FITS}}$	& \multicolumn{4}{c}{20}  \\
$N_{\rm{D-FITS}}$	& \multicolumn{4}{c}{20}  \\
$N_{\rm{STEP}}$		& --			& \multicolumn{3}{c}{1024} \\
$N_{\rm{SCAN}}$		& --			& \multicolumn{3}{c}{150} \\
$L_{\rm{SCAN}}$		& --			& \multicolumn{3}{c}{140$\mu$m} \\
$N_{\rm{MED}}$		& \multicolumn{4}{c}{3 (air, BK7, F7)} \\
$\mats{\zeta}$		& \multicolumn{4}{c}{See Fig.~\ref{fig:tc_pavosim_zeta_D}} \\
$N_{\rm{TEL}}$		& \multicolumn{4}{c}{2 (N3, S1)} \\
$\mata{B}$		& \multicolumn{4}{c}{$|B_{1,2}|\sim$15m} \\
\cmidrule{2-5}
\multirow{3}{*}{$m_V$}	& \multicolumn{4}{c}{Sub-case A: 0.0} \\
			& \multicolumn{4}{c}{Sub-case B: 2.0} \\
			& \multicolumn{4}{c}{Sub-case C: 4.0} \\
\cmidrule{2-5}
$\mata{V}^2$		& 0.01--0.36 & 0.25 & 0.10 & 0.01 \\
$\mata{D}, \mata{d}$	& \multicolumn{4}{c}{See Fig.~\ref{fig:tc_pavosim_zeta_D}} \\
$r_0$			& \multicolumn{4}{c}{10cm} \\
$\sigma_{r_0}^{-1}$	& \multicolumn{4}{c}{0.5$\mu$m} \\
$\tau_0$		& \multicolumn{4}{c}{1ms} \\
$L_0$			& \multicolumn{4}{c}{100m} \\
\hline
\bottomrule
\end{tabular}
\end{table}

%% file: main.bbl
\begin{thebibliography}{21}
\providecommand{\natexlab}[1]{#1}
\providecommand{\url}[1]{{#1}}
\providecommand{\urlprefix}{URL }
\expandafter\ifx\csname urlstyle\endcsname\relax
  \providecommand{\doi}[1]{DOI~\discretionary{}{}{}#1}\else
  \providecommand{\doi}{DOI~\discretionary{}{}{}\begingroup
  \urlstyle{rm}\Url}\fi
\providecommand{\eprint}[2][]{\url{#2}}

\bibitem[{{Bessell}(1979)}]{Bessell:1979}
{Bessell} MS (1979) {UBVRI photometry. II - The Cousins VRI system, its
  temperature and absolute flux calibration, and relevance for two-dimensional
  photometry}. \pasp\ 91:589--607, \doi{10.1086/130542}

\bibitem[{Colavita et~al(1999)Colavita, Wallace, Hines, Gursel, Malbet, Palmer,
  Pan, Shao, Yu, Boden, and Others}]{Colavita:1999}
Colavita M, Wallace J, Hines B, Gursel Y, Malbet F, Palmer D, Pan X, Shao M, Yu
  J, Boden A, Others (1999) {The Palomar testbed interferometer}. \apj\
  510:505--521, \doi{10.1086/306579}

\bibitem[{{Colavita}(1992)}]{Colavita:1992}
{Colavita} MM (1992) {Phase Referencing for Stellar Interferometry}. In:
  {Beckers} JM, {Merkle} F (eds) European Southern Observatory Conference and
  Workshop Proceedings, European Southern Observatory Conference and Workshop
  Proceedings, vol~39, pp 845--851

\bibitem[{{Davis} et~al(1999){Davis}, {Tango}, {Booth}, {ten Brummelaar},
  {Minard}, and {Owens}}]{Davis:1999}
{Davis} J, {Tango} WJ, {Booth} AJ, {ten Brummelaar} TA, {Minard} RA, {Owens} SM
  (1999) {The Sydney University Stellar Interferometer - I. The instrument}.
  \mnras\ 303:773--782, \doi{10.1046/j.1365-8711.1999.02269.x}

\bibitem[{{Glindemann}(2011)}]{Glindemann:2011}
{Glindemann} A (2011) {Principles of Stellar Interferometry}. Astronomy and
  Astrophysics Library, Springer

\bibitem[{{Ireland} et~al(2008){Ireland}, {M{\'e}rand}, {ten Brummelaar},
  {Tuthill}, {Schaefer}, {Turner}, {Sturmann}, {Sturmann}, and
  {McAlister}}]{Ireland:2008}
{Ireland} MJ, {M{\'e}rand} A, {ten Brummelaar} TA, {Tuthill} PG, {Schaefer} GH,
  {Turner} NH, {Sturmann} J, {Sturmann} L, {McAlister} HA (2008) {Sensitive
  visible interferometry with PAVO}. In: \procspie\, vol 7013,
  \doi{10.1117/12.788386}

\bibitem[{{Kok} et~al(2012){Kok}, {Ireland}, {Tuthill}, {Robertson},
  {Warrington}, and {Tango}}]{Kok:2012}
{Kok} Y, {Ireland} MJ, {Tuthill} PG, {Robertson} JG, {Warrington} BA, {Tango}
  WJ (2012) {Self-phase-referencing interferometry with SUSI}. In: \procspie\,
  vol 8445, \doi{10.1117/12.925238}

\bibitem[{{Lane} and {Muterspaugh}(2004)}]{Lane:2004}
{Lane} BF, {Muterspaugh} MW (2004) {Differential Astrometry of Subarcsecond
  Scale Binaries at the Palomar Testbed Interferometer}. \apj\ 601:1129--1135,
  \doi{10.1086/380760}

\bibitem[{{Lane} et~al(1992){Lane}, {Glindemann}, and {Dainty}}]{Lane:1992}
{Lane} RG, {Glindemann} A, {Dainty} JC (1992) {Simulation of a Kolmogorov phase
  screen}. Waves in Random Media 2:209--224, \doi{10.1088/0959-7174/2/3/003}

\bibitem[{{Lawson} et~al(2000){Lawson}, {Colavita}, {Dumont}, and
  {Lane}}]{Lawson:2000}
{Lawson} PR, {Colavita} MM, {Dumont} PJ, {Lane} BF (2000) {Least-squares
  estimation and group delay in astrometric interferometers}. In: \procspie\,
  vol 4006, pp 397--406

\bibitem[{{McAlister} et~al(2005){McAlister}, {ten Brummelaar}, {Gies},
  {Huang}, {Bagnuolo}, {Shure}, {Sturmann}, {Sturmann}, {Turner}, {Taylor},
  {Berger}, {Baines}, {Grundstrom}, {Ogden}, {Ridgway}, and {van
  Belle}}]{McAlister:2005}
{McAlister} HA, {ten Brummelaar} TA, {Gies} DR, {Huang} W, {Bagnuolo} WG Jr,
  {Shure} MA, {Sturmann} J, {Sturmann} L, {Turner} NH, {Taylor} SF, {Berger}
  DH, {Baines} EK, {Grundstrom} E, {Ogden} C, {Ridgway} ST, {van Belle} G
  (2005) {First Results from the CHARA Array. I. An Interferometric and
  Spectroscopic Study of the Fast Rotator {$\alpha$} Leonis (Regulus)}. \apj\
  628:439--452, \doi{10.1086/430730}

\bibitem[{{McGlamery}(1976)}]{McGlamery:1976}
{McGlamery} BL (1976) {Computer simulation studies of compensation of
  turbulence degraded images}. In: \procspie\, vol~74, pp 225--233

\bibitem[{{Muterspaugh} et~al(2010){Muterspaugh}, {Lane}, {Kulkarni},
  {Konacki}, {Burke}, {Colavita}, {Shao}, {Wiktorowicz}, and
  {O'Connell}}]{Muterspaugh:2010a}
{Muterspaugh} MW, {Lane} BF, {Kulkarni} SR, {Konacki} M, {Burke} BF, {Colavita}
  MM, {Shao} M, {Wiktorowicz} SJ, {O'Connell} J (2010) {The Phases Differential
  Astrometry Data Archive. I. Measurements and Description}. \aj\
  140:1579--1622, \doi{10.1088/0004-6256/140/6/1579}, \eprint{1010.4041}

\bibitem[{{Robertson} et~al(2010){Robertson}, {Ireland}, {Tango}, {Davis},
  {Tuthill}, {Jacob}, {Kok}, and {Ten Brummelaar}}]{Robertson:2010}
{Robertson} JG, {Ireland} MJ, {Tango} WJ, {Davis} J, {Tuthill} PG, {Jacob} AP,
  {Kok} Y, {Ten Brummelaar} TA (2010) {Instrumental developments for the Sydney
  University Stellar Interferometer}. In: \procspie\, vol 7734,
  \doi{10.1117/12.856557}

\bibitem[{{Roddier}(1981)}]{Roddier:1981}
{Roddier} F (1981) {The Effects of Atmospheric Turbulence in Optical
  Astronomy}. \progopt\ 19:281--376, \doi{10.1016/S0079-6638(08)70204-X}

\bibitem[{{Sasiela} and {Shelton}(1993)}]{Sasiela:1993}
{Sasiela} RJ, {Shelton} JD (1993) {Transverse spectral filtering and Mellin
  transform techniques applied to the effect of outer scale on tilt and tilt
  anisoplanatism}. J~Opt~Soc~Am~A 10:646--660, \doi{10.1364/JOSAA.10.000646}

\bibitem[{{Shaklan}(1989)}]{Shaklan:1989}
{Shaklan} SB (1989) {Multiple Beam Correlation Using Single-Mode Fiber Optics
  with Application to Interferometric Imaging}. PhD thesis, The University of
  Arizona

\bibitem[{Shao and Colavita(1992)}]{Shao:1992}
Shao M, Colavita M (1992) {Potential of long-baseline infrared interferometry
  for narrow-angle astrometry}. \aap\ 262:353--358

\bibitem[{{Tango}(1990)}]{Tango:1990}
{Tango} WJ (1990) {Dispersion in stellar interferometry}. \ao\ 29:516--521,
  \doi{10.1364/AO.29.000516}

\bibitem[{{Taylor}(1938)}]{Taylor:1938}
{Taylor} GI (1938) {The Spectrum of Turbulence}. Royal Society of London
  Proceedings Series A 164:476--490, \doi{10.1098/rspa.1938.0032}

\bibitem[{{Valley}(1979)}]{Valley:1979}
{Valley} GC (1979) {Long- and short-term Strehl ratios for turbulence with
  finite inner and outer scales}. \ao\ 18:984--987, \doi{10.1364/AO.18.000984}

\end{thebibliography}
